\documentclass[12pt]{article}

\usepackage{amsmath,amsfonts,amssymb,amsthm}
\usepackage{algorithm}
\usepackage{algpseudocode}
\algnewcommand\algorithmicforeach{\textbf{for each}}
\algdef{S}[FOR]{ForEach}[1]{\algorithmicforeach\ #1\ \algorithmicdo}
\usepackage{subcaption}

\makeatletter
\def\algbackskip{\hskip-\ALG@thistlm}
\makeatother

\usepackage{bbding}
\usepackage{graphicx,psfrag,epsf}
\usepackage{enumitem}
\usepackage{bm}
\usepackage{bbm}
\usepackage{soul} % underline
\usepackage[framemethod=TikZ]{mdframed}

\usepackage{natbib}

\usepackage{xurl} % not crucial - just used below for the URL
\usepackage[colorlinks,allcolors=blue]{hyperref}

\usepackage{comment}

\usepackage{algorithm,algpseudocode}

\makeatletter
\newenvironment{breakablealgorithm}
  {% \begin{breakablealgorithm}
   \begin{center}
     \refstepcounter{algorithm}% New algorithm
     \hrule height.8pt depth0pt \kern2pt% \@fs@pre for \@fs@ruled
     \renewcommand{\caption}[2][\relax]{% Make a new \caption
       {\raggedright\textbf{\fname@algorithm~\thealgorithm} ##2\par}%
       \ifx\relax##1\relax % #1 is \relax
         \addcontentsline{loa}{algorithm}{\protect\numberline{\thealgorithm}##2}%
       \else % #1 is not \relax
         \addcontentsline{loa}{algorithm}{\protect\numberline{\thealgorithm}##1}%
       \fi
       \kern2pt\hrule\kern2pt
     }
  }{% \end{breakablealgorithm}
     \kern2pt\hrule\relax% \@fs@post for \@fs@ruled
   \end{center}
  }
\makeatother

\usepackage[section]{placeins}
\usepackage{rotating}
\usepackage{placeins}
\usepackage{setspace}
\usepackage{relsize}
\usepackage{scalerel}

\usepackage{tikz}
\usetikzlibrary{tikzmark, arrows, arrows.meta, bending, calc, shapes, patterns, decorations, decorations.pathreplacing, decorations.markings, fit, positioning, matrix}

\usepackage{epigraph}
\newcommand{\blind}{0}

\newcommand{\bstar}{\mathsmaller{\text{\FiveStar}}}
\newcommand{\bA}{\bm{A}}
\newcommand{\ba}{\bm{a}}
\newcommand{\bE}{\bm{E}}

\newcommand{\bp}{\bm{p}}
\newcommand{\bU}{\bm{U}}
\newcommand{\bu}{\bm{u}}
\newcommand{\bV}{\bm{V}}
\newcommand{\bv}{\bm{v}}
\newcommand{\bW}{\bm{W}}

\newcommand{\bpa}{\bm{pa}}

\newcommand{\cA}{\mathcal{A}}
\newcommand{\cB}{\mathcal{B}}
\newcommand{\cD}{\mathcal{D}}
\newcommand{\cC}{\mathcal{C}}
\newcommand{\cE}{\mathcal{E}}
\newcommand{\cF}{\mathcal{F}}
\newcommand{\cG}{\mathcal{G}}

\newcommand{\cP}{\mathcal{P}}

\newcommand{\cS}{\mathcal{S}}
\newcommand{\cT}{\mathcal{T}}
\newcommand{\cU}{\mathcal{U}}
\newcommand{\E}{\mathbb{E}}

\newcommand{\bbone}{\mathbbm{1}}

\DeclareMathOperator*{\argext}{arg\,extremum}
\newcommand\hash{\scalebox{0.8}{\raisebox{0.4ex}{\#}}} % fix baseline

\renewcommand{\langle}{\{}
\renewcommand{\rangle}{\}}

\renewcommand{\l}{\left}
\renewcommand{\r}{\right}

\colorlet{editcolor}{purple!90!black}

\setstcolor{editcolor}

\newtheorem{theorem}{Theorem}
\newtheorem{proposition}{Proposition}

\newtheorem{corollary}{Corollary}%[lemma]

\begin{document}

\widowpenalty=0
\clubpenalty=0
%\flushbottom

\def\spacingset#1{\renewcommand{\baselinestretch}%
{#1}\small\normalsize} \spacingset{1}

\usetikzlibrary{shapes,shapes.geometric, decorations,arrows,calc,arrows.meta,fit,positioning}
\tikzset{
 -Latex,auto,node distance=1 cm and 1 cm,semithick,
 state/.style={ellipse, draw, minimum width=0.7 cm},
 point/.style={circle, draw, inner sep=0.04cm,fill,node contents={}},
 bidirected/.style={Latex-Latex,dashed},
 el/.style={inner sep=2pt, align=left, sloped}
}

%%%%%%%%%%%%%%%%%%%%%%%%%%%%%%%%%%%%%%%%%%%%%%%%%%%%%%%%%%%%%%%%%%%%%%%%%%%%%%

 \if0\blind
{
 \title{ \vspace{-.6in} \bf An Automated Approach to\\Causal Inference in Discrete Settings\thanks{\scriptsize
 Guilherme Jardim Duarte is a Ph.D.\ student in the Operations, Information and Decisions Department, the Wharton School of the University of Pennsylvania. Noam Finkelstein is a Ph.D.\ student in the Department of Computer Science, Johns Hopkins University. Dean Knox is an Andrew Carnegie Fellow and an assistant professor in the Operations, Information and Decisions Department, the Wharton School of the University of Pennsylvania. Jonathan Mummolo is an assistant professor of Politics and Public Affairs, Princeton University. Ilya Shpitser is the John C.\ Malone Assistant Professor in the Department of Computer Science, Whiting School of Engineering at the Johns Hopkins University. Authors listed in alphabetical order. For helpful feedback, we thank Peter Aronow, Justin Grimmer, Kosuke Imai, Luke Keele, Gary King, Christopher Lucas, Fredrik S\"avje, Brandon Stewart, Eric Tchetgen Tchetgen, and participants in the Harvard Applied Statistics Workshop, the New York University Data Science Seminar, University of Pennsylvania Causal Inference Seminar, PolMeth 2021, and the Yale Quantitative Research Methods Workshop. We gratefully acknowledge financial support from AI for Business and the Analytics at Wharton Data Science and Business Analytics Fund. This research was made possible in part by a grant from the Carnegie Corporation of New York. The statements made and views expressed are solely the responsibility of the authors.
 }}
 \author{
 Guilherme Duarte\\
 \href{mailto:gjduarte@upenn.edu}{\tt gjduarte@upenn.edu}
 \and
 Noam Finkelstein\\
 \href{mailto:noam@jhu.edu}{\tt noam@jhu.edu}
 \and
 Dean Knox\\
 \href{mailto:dcknox@upenn.edu}{\tt dcknox@upenn.edu}
 \and
 Jonathan Mummolo\\
 \href{mailto:jmummolo@princeton.edu}{\tt jmummolo@princeton.edu}
 \and
 Ilya Shpitser\\
 \href{mailto:ilyas@cs.jhu.edu}{\tt ilyas@cs.jhu.edu}
 }
 \date{First draft: February 10, 2021\\This draft: \today}
 \maketitle
} \fi

\if1\blind
{\begin{center}
 {\LARGE\bf \bf An Automated Approach to\\[1ex]Causal Inference in Discrete Settings}
\end{center}
 \medskip
} \fi

\thispagestyle{empty}
%\clearpage

\vspace{-.25in}
\begin{abstract}
When causal quantities cannot be point identified, researchers often pursue partial identification to quantify the range of possible values. However, the peculiarities of applied research conditions can make this analytically intractable. We present a general and automated approach to causal inference in discrete settings. We show causal questions with discrete data reduce to polynomial programming problems, and we present an algorithm to automatically bound causal effects using efficient dual relaxation and spatial branch-and-bound techniques. The user declares an estimand, states assumptions, and provides data (however incomplete or mismeasured). The algorithm then searches over admissible data-generating processes and outputs the most precise possible range consistent with available information---i.e., \emph{sharp} bounds---including a point-identified solution if one exists. Because this search can be computationally intensive, our procedure reports and continually refines non-sharp ranges that are guaranteed to contain the truth at all times, even when the algorithm is not run to completion. Moreover, it offers an additional guarantee we refer to as $\varepsilon$-sharpness, characterizing the worst-case looseness of the incomplete bounds. Analytically validated simulations show the algorithm accommodates classic obstacles, including confounding, selection, measurement error, noncompliance, and nonresponse. 
\end{abstract}

%\begin{comment}
\noindent%
\vfill
\thispagestyle{empty}
%\end{comment}
\newpage

\tableofcontents
\thispagestyle{empty}

\newpage
 \setcounter{page}{1}
\epigraph{The combination of some data and an aching desire for an answer does not ensure that a reasonable answer can be extracted from a given body of data.}{\citet[][pp.\ 74--75]{tukey1986}}

%\spacingset{1.9} % DON'T change the spacing!
\spacingset{1.1} % DON'T tell me what to do

%\vspace{-.6cm}
\section{Introduction}
\label{sec:intro} 
%\vspace{-.4cm}

When causal quantities cannot be point identified, researchers often pursue partial identification to quantify the range of possible answers. These solutions are tailored to specific scenarios \citep[e.g.][]{Lee2009,garbiel2020,kennedy2019survivor,KnoxLoweMummolo2020,LiPearl2021_Bounds,Sjolander2014}, but the idiosyncrasies of applied research can render prior results unusable if identifying assumptions fail or slightly differing causal structures are encountered. This case-by-case approach to deriving causal bounds presents a major obstacle to scientific progress. To increase the pace of discovery, researchers need a general approach that is robust to context-specific peculiarities.

%We first prove causal questions with discrete data reduce to polynomial programming problems, then present an algorithm to sharply bound causal effects using efficient dual relaxation and spatial branch-and-bound techniques \citep{vigerske2018scip,gamrath2020scip}.

In this paper, we present an automated approach to causal inference in discrete settings which can be applied to all graphical causal models, as well as all observed quantities and domain assumptions in standard use. With our algorithm, users declare an estimand, state assumptions, and provide available data---however incomplete or mismeasured. The algorithm then outputs \emph{sharp bounds}, the most precise possible answer to the causal query given these inputs, including
a point estimate if the solution is identified. This approach can accommodate scenarios involving any classic threat to inference, including but not limited to missing data, selection, measurement error, and noncompliance. Our algorithm also has the desirable property of alerting users when assumptions conflict with observed data, indicating a faulty causal theory. Finally, we develop techniques for drawing statistical inferences about estimated bounds. We demonstrate our method using a host of simulations, validating results wherever existing analytic solutions are available.

Our work advances a rich literature on partial identification in causal inference \citep{robins1989,manski1990,Heckman2001,ZhangRubin2003,cai2008,swanson2018,garbiel2020,Molinari2020}, outlined in Section~\ref{sec:lit}, which has sometimes cast the task as a constrained optimization problem that can be solved computationally. In pioneering work, \citeauthor{balke1994counterfactual} (\citeyear{balke1994counterfactual}, \citeyear{balke1997bounds}) provided a method for calculating sharp bounds when causal queries can be expressed as linear programming problems. However, a wide range of estimands and empirical obstacles result in causal queries that are \emph{not} reducible to linear programs, and a complete computational solution has remained elusive.

When feasible, sharp-bounding approaches offer a principled and transparent approach to causal inference that makes maximum use of available information while acknowledging its limitations. Claims outside the bounds can be immediately rejected, and claims inside the bounds must be explicitly justified by additional assumptions or data that enable tightening. But several obstacles still preclude widespread use of these techniques. For one, analytic derivation remains intractable for many problems. Within the subclass of linear problems, \citeauthor{balke1994counterfactual}'s (\citeyear{balke1994counterfactual}) simplex method offers a highly efficient analytic solution, but one that fails to generalize to the many partially observed settings where nonlinearity arises. Analytic nonlinear solutions remain limited to specific results, painstakingly derived case by case \citep[e.g.][]{KnoxLoweMummolo2020,garbiel2020,LiPearl2021_Bounds}. Though general sharp bounds can in theory be obtained by various nonlinear optimization techniques \citep{geiger1999quantifier,zhang2021pi}, such approaches are often computationally infeasible. This is because without exhaustively exploring a vast model space, analysts can obtain local optima that correspond to potentially invalid bounds---i.e., ranges that may fail to contain the truth. 

%Taken together, computational approaches for discovering causal bounds remain a promising but practically limited avenue for applied research.  

To address these limitations, we first show in Sections~\ref{sec:prelim} and \ref{sec:poly} that essentially all common causal queries involving discrete variables can be reduced to polynomial programs---a well-studied class of optimization tasks that nest linear programming as a special case---building on prior results from \citet{geiger1999quantifier} and \citet{wolfe2019inflation}.\footnote{Specifically, our results apply to elementary arithmetic functionals or monotonic transformations thereof---a broad set that essentially includes all causal assumptions, observed quantities, and estimands in standard use. For example, the average treatment effect and the log odds ratio can be sharply bounded with our approach, but non-analytic functionals (which are rarely if ever encountered) cannot. Functionals that do not meet these conditions can be approximated to arbitrary precision, if they have convergent power series.} While mature techniques have been developed for such tasks \citep{belelimawa08,vigerske2018scip,gamrath2020scip}, it is well known that solving polynomial programs to global optimality is in general NP-hard. The difficulty of the problem thus highlights the need for efficient algorithms and bounding techniques that remain valid even when analysts are faced with time constraints. In Sections~\ref{sec:simp}--\ref{sec:algo}, we develop a procedure, based on dual relaxation and spatial branch-and-bound relaxation techniques, that provides valid bounds of arbitrary sharpness, for all causal structures, under virtually any information environments and domain assumptions. We show this procedure is guaranteed to achieve complete sharpness with sufficient computation time; in smaller problems, this can occur in a matter of seconds. However, in cases where the time needed to discover sharp bounds is prohibitive---which can occur even in moderately sized problems with severe information fragmentation---our algorithm is \emph{anytime} \citep{DeanBoddy1988}, meaning it can be interrupted to obtain non-sharp bounds that are nonetheless guaranteed to be valid. Our technique also offers an additional guarantee we term ``$\varepsilon$-sharpness," indicating the worst-case looseness factor of the relaxed bounds relative to the unknown, completely sharp bounds. In Section~\ref{sec:inference}, we provide two approaches for characterizing uncertainty in the estimated bounds, and we demonstrate our technique in a series of simulations in Section~\ref{sec:sims}. Our simulations, validated against previously derived analytically results where possible, show the flexibility of our approach and the ease with which assumptions can be modularly imposed or relaxed. Moreover, we demonstrate how the algorithm can uncover counterintuitive results: in one case, we show a scenario that appears to be partially identified is in fact point identified, improving over widely used bounds \citep{manski1990} and recovering a recent advance in the literature on nonrandom missingness \citep{miao2015identification}. 

Our approach offers a complete and computationally feasible approach to causal inference in discrete settings. Given a well-defined causal query, valid assumptions, and data, researchers now have a general and automated process to draw causal inferences that are guaranteed to be valid and, with sufficient computation time, provably optimal. As we discuss in Sections~\ref{sec:critique}--\ref{sec:discussion}, our approach's modular nature also allows analysts to conduct principled robustness tests and sensitivity analyses that can identify the most promising avenues for future research, promote research transparency, and accelerate scientific discovery.

%Our approach offers a complete and computationally feasible approach to causal inference in discrete settings. Given a well-defined causal query, valid assumptions, and data, researchers now have a general and automated process to draw causal inferences that are guaranteed to be valid and, with sufficient computation time, provably optimal. As we discuss in  

%\vspace{-.6cm}
\section{Related Literature}\label{sec:lit}
%\vspace{-.2cm}

Researchers have long
sought to automate causal identification by recasting causal queries as constrained
optimization problems that can be solved computationally. Our work is most closely related to \citet{balke1994counterfactual,balke1997bounds}, which
showed that certain bounding problems in discrete settings---generally corresponding to causal systems in which outcomes and manipulated variables are fully observed---could be formulated
as the minimization and maximization of a linear objective function subject to
linear equality and inequality constraints. In these cases, causal bounding problems can be reformulated as linear programming problems,
which admit both symbolic solutions and highly efficient numerical solutions. Subsequent studies have proven that in particular settings, the bounds produced by this technique are sharp \citep{Ramsahai2012,bonet2001_instrumentality,Heckman2001}, and \cite{sachs2020} shows this approach produces sharp bounds for any such linear problem. These results were extended by \citet{geiger1999quantifier}, which showed that a
much broader class of discrete problems can be formulated in terms of polynomial relations---at least, when analysts have precise information about the kinds of disturbances or confounders that may exist, expressed in terms of latent variable cardinalities. These discrete problems include not only the bounds studied in this
paper, but the related problem of determining what constraints on the main
variables are implied by a causal graph. In addition to the well-known
conditional independence constraints implied by d-separation,
these can include generalized equality constraints \citep[or Verma
constraints;][]{verma1990_equivalence,tian2002_testable}. Beyond these
equalities, the main variables are also constrained by generalizations of the
instrumental inequalities
\citep{pearl1995testability,bonet2001_instrumentality}.

\citet{geiger1999quantifier} note that in theory, algorithms for quantifier
elimination can provide symbolic solutions for these questions. However, the
time complexity of quantifier elimination is doubly exponential, rendering it
infeasible for all but the simplest cases. At the core of this issue is that
symbolic methods provide a general solution, meaning that they must explore the
space of all possible inputs. In contrast, numerical methods such as our approach can often eliminate
portions of the space that are irrelevant, accelerating computation.% In the following section, we develop numerical polynomial programming techniques to solve any discrete bounding problem.

Even so, computation can be time-consuming; polynomial programming is in general
NP-hard. In practice, many optimizers are able to rapidly find reasonably good
values but cannot guarantee optimality without exhaustively searching the space 
of candidates. This approach poses a challenge for obtaining causal bounds, which
represent minimal and maximal values of the estimand under all models that are \textit{admissible}, or consistent with observed data and
modeling assumptions. If a local optimizer operates on the original problem (the \textit{primal}), proceeding from the interior and widening bounds as more extreme models are discovered, then failing to reach global optimality will result in \textit{invalid bounds}---ranges narrower
than the optimal sharp bounds which do not contain all possible solutions. 

In this paper, we detail an approach that resolves this obstacle by allowing analysts to obtain valid bounds in limited time. At a high level, our approach is to reexpress causal inference problems in terms of principal strata \citep{Frangakis2002}. To do so, we first present new results on lossless reductions for latent variables of unknown cardinality. We then show that causal estimands, modeling assumptions, and observed information can all be expressed in terms of polynomial expressions, equalities, and inequalities with no loss of information. We show how these systems can be simplified for computational efficiency, then develop an iterative primal-dual algorithm that searches for admissible models from the interior of the bounds (the primal problem) while simultaneously refining a guaranteed-valid outer envelope for the sharp bounds (the dual problem). Even when exhaustive search is computationally infeasible, suboptimal primal and dual values can still be found and improved over time. We show suboptimal dual points allow analysts to report valid \textit{loose bounds}---those that are wider than the unknown sharp bounds. Our method also utilizes the suboptimal primal points, allowing analysts to assess the worst-case \textit{looseness factor} of the reported valid bounds, relative to the unknown sharp bounds.

%\vspace{-.6cm}
\section{Preliminaries}\label{sec:prelim}
%\vspace{-.2cm}

In this section, we define notation and discuss concepts necessary to derive our key results. We first review how arbitrary directed acyclic graphs (DAGs) can be ``canonicalized'' without loss of information, resulting in an equivalent form with properties amenable to analysis \citep{evans2018margins}. We then describe how graphs in this form give rise to potential outcomes and principal strata \citep{Frangakis2002}, two key building blocks in our analytic strategy. 

Suppose that for each i.i.d.\ unit $i \in \langle 1, \ldots, N \rangle$, the
main variables of interest are contained in $\bV_i = \langle V_{i,1}, \ldots,
V_{i,J} \rangle$, indexed by $j$. We will suppose that the sample space of each
main variable, $\cS(V_{i,j})$, has finite cardinality. These variables may be
either observed or unobserved---as we will show, it is often useful to
reason about unobserved elements of $\bV_i$ in the context of missing data and
measurement error. We will also consider unobserved causal ancestors of $\bV_i$, collectively denoted $\bU_i = \langle U_{i,1}, \ldots, U_{i,K}\rangle$ and indexed by $k$, that represent random disturbances or confounders. Without loss of generality, these disturbances---which have unknown, possibly infinite cardinality---are assumed to subsume all phenomena that are causally relevant to $\bV_i$.\footnote{We note that traditionally, variables in $\bV_i$ are permitted to be affected by exogenous causal noise not represented in the graph. By incorporating all causally relevant factors into $\bU_i$, we take each variable in $\bV_i$ to be a deterministic function of its parents in the graph, discussed in more detail below.} As we show in Section~\ref{sec:po_strata}, this assumption is without consequence, because even a continuous and infinite dimensional $\bU_i$ must still map down to the same finite canonical partitions that we describe there. In addition, we will make use of \emph{counterfactual} random variables, which represent hypothetical versions of random variables in $\bV_i$ that would have occurred had, contrary to fact, treatment variables been exogenously set to a specified value. (A more rigorous definition is given in Section~\ref{sec:po}.) By convention, bold letters denote collections of variables;
uppercase and lowercase letters respectively denote random
variables and their realizations. We will consider population distributions
until discussing inference in Section~\ref{sec:inference}.

%\vspace{-.3cm}
\subsection{Canonical DAGs}
\label{sec:canon}

We now discuss how DAGs can be canonicalized, or distilled to minimal form, to clarify which aspects of the structural model can be ignored, greatly simplifying the bounding task. Suppose that causal relationships between all variables in $\bV_i$ and $\bU_i$ are represented by a directed acyclic graph (DAG) $\cG$.
The nonparametric
structural equations model (NPSEM) of a DAG states that each main variable $V_{i,j} \in \bV_i$ is a
deterministic function of its parents in the graph $\cG$, denoted $\bpa(V_{i,j})$. That is, all factors determining $V_{i,j}$ are contained in $\bpa(V_{i,j})$, a subset of $\bU_i$ and $\bV_i$. We denote the function mapping from $\bpa(V_{i,j})$ to $V_{i,j}$ as $V_{i, j} = f_j\big({\bpa}(V_{i,j})\big)$; we use $\cF = \langle f_1, \ldots, f_J
\rangle$ to denote the collection of these structural equations, or the \textit{structural causal model model} of $\bV$ \citep{pearl2009causality, richardson2013singlewi}. Note that each main variable may be influenced by
multiple disturbances, and a single disturbance may influence multiple main
variables.

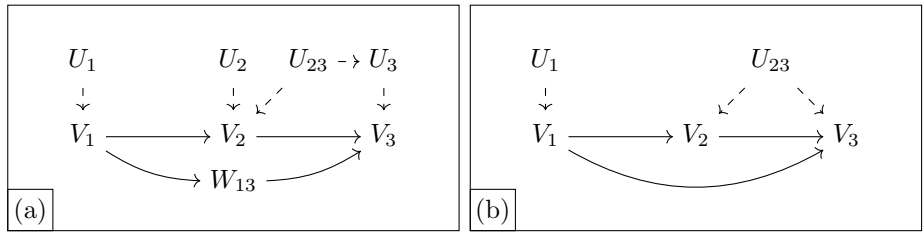
\begin{figure}
  \caption{ \textbf{Canonicalization of a mediation model.} Mediation DAG in
    non-canonical form (panel a) and canonical form (panel b) that are fully
    equivalent with respect to their full data law. Unit indices, $i$, are
    suppressed. Canonicalization proceeds as follows: (i) the dependent
    disturbance $U_3$ is absorbed into its parent $U_{23}$; (ii) the superfluous
    $U_2$ is eliminated as it influences a subset of $U_{23}$'s children; and
    (iii) the irrelevant $W_{13}$ is absorbed into the $V_1 \rightarrow V_3$
    path as it is neither observed nor of interest. A complete guide to
    canonicalization is given in Appendix~\ref{app:canonical_dag}.
    }
  \label{fig:mediation}
  \centering
%  \vspace{.5ex}%
    \begin{minipage}[t]{.37\textwidth}
      \footnotesize
      \centering
      \begin{tikzpicture}[text height=2ex, text depth=1ex]
        %% \node (W0) at (-1,0) {$W_0$};
        \node (W13) at (2,-.6) {$W_{13}$};
        \node (U1) at (0,1) {$U_1$};
        \node (U2) at (2,1) {$U_2$};
        \node (U23) at (3,1) {$U_{23}$};
        \node (U3) at (4,1) {$U_3$};
        \node (V1) at (0,0) {$V_1$};
        \node (V2) at (2,0) {$V_2$};
        \node (V3) at (4,0) {$V_3$};
        \path[dashed] (U1) edge [->](V1);
        \path[dashed] (U2) edge [->](V2);
        \path[dashed] (U23) edge [->](V2);
        \path[dashed] (U23) edge [->](U3);
        \path[dashed] (U3) edge [->](V3);
        %% \path (W0) edge [->](V1);
        \path (V1) edge [->](V2);
        \path (V2) edge [->](V3);
        \path (V1) edge [->, bend right=15](W13);
        \path (W13) edge [->, bend right=15](V3);
        %% \draw (-.5, -1) rectangle (4.5, 1.5);
        \useasboundingbox (-1, -1.25) rectangle (5, 1.75);
        \node (a) at (-.7, -1) {(a)};
        \draw (-1, -1.25) rectangle (-.4, -.7);
        \draw (-1, -1.25) rectangle (5, 1.75);
      \end{tikzpicture}
      \\
    \end{minipage}
  \begin{minipage}[t]{.37\textwidth}
    \footnotesize
    \centering
      \begin{tikzpicture}[text height=2ex, text depth=1ex]
        \node (U1) at (0,1) {$U_1$};
        %% \node (U2) at (2,1) {$U_2$};
        \node (U23) at (3,1) {$U_{23}$};
        %% \node (U3) at (4,1) {$U_3$};
        \node (V1) at (0,0) {$V_1$};
        \node (V2) at (2,0) {$V_2$};
        \node (V3) at (4,0) {$V_3$};
        \path[dashed] (U1) edge [->](V1);
        %% \path[dashed] (U2) edge [->](V2);
        \path[dashed] (U23) edge [->](V2);
        \path[dashed] (U23) edge [->](V3);
        %% \path[dashed] (U3) edge [->](V3);
        \path (V1) edge [->](V2);
        \path (V2) edge [->](V3);
        \path (V1) edge [->, bend right=30](V3);
        %% \draw (-.5, -1) rectangle (4.5, 1.5);
        \useasboundingbox (-1, -1.25) rectangle (5, 1.75);
        \node (b) at (-.7, -1) {(b)};
        \draw (-1, -1.25) rectangle (-.4, -.7);
        \draw (-1, -1.25) rectangle (5, 1.75);
      \end{tikzpicture}
      \\
  \end{minipage}
\end{figure}

A DAG is said to be in canonical form if (i) all disturbances are
exogeneous, i.e.\ no variable in $\bU_i$ has any parents in $\cG$; and (ii)
there exists no pair of disturbances, $U_i$ and $U'_i$, such that $U_i$ influences a subset of the
variables influenced by $U'_i$. \citet{evans2018margins} showed that for any DAG $\cG'$ not in canonical form;
there exists a canonical form DAG $\cG$ with an identical distribution over all factual and
counterfactual versions of all variables in $\bV_i$. We can therefore without loss of generality limit our consideration to DAGs in canonical form.
An example of a DAG not in canonical form is given in panel
Figure~\ref{fig:mediation}(a).
Panel Figure~\ref{fig:mediation}(b) illustrates the canonicalized version of this
graph. For convenience, we will refer to the joint distribution over all factual and counterfactual versions of $\bV_i$ as the \emph{full data law}. Moreover, any DAG over $\bU_i$, $\bV_i$, and unobserved ancillary
variables $\bW_i$ with unknown cardinality (e.g., confounders or mediators not of direct interest) also has an equivalent canonical DAG with respect to
this full data law. A guide for canonicalizing arbitrary DAGs is given in
Appendix~\ref{app:canonical_dag}.

In short, representing the causal graph in canonical form distills the data-generating process (DGP) to its simplest form, eliminating potentially complex networks of disturbances. Removing variables that are irrelevant to the causal goal further simplifies the structure. Without these simplifications, it would be exceedingly difficult, if not intractable, to convert causal problems into polynomial programs that can be readily optimized---the essence of the approach we develop below. 

%\vspace{-.3cm}
\subsection{Potential Outcomes}
\label{sec:po}

The notation of potential outcome functions allows us to compactly
express the effects of manipulating a variable's parents or other ancestors. Let $\bA \subset \bV$ be a subset of variables that will be intervened
upon, fixing them to $\bA = \ba$. When $\bA = \varnothing$, so that no
intervention occurs, then define $V_{i,j}(\ba) = V_{i,j}$, the natural
value. When $\bA \subseteq \bpa(V_{i,j})$, so that only immediate parents are
manipulated, then unit $i$'s potential outcome function is given by its response
function, $V_{i,j}(\ba) = f_j\big(\bA = \ba, \bpa(V_{i,j}) \setminus \bA\big)$. We will now
define more general potential outcome functions by \emph{recursive substitution}
\citep{richardson2013singlewi,Shpitser2018_id_gcm}. For arbitrary interventions on $\bA \subset \bV$,
let $V_{i,j}(\ba) = V_{i,j}\big( \{ a_{\ell} : \bA_{\ell} \in \bpa(V_{i,j}) \} \cup
\{ V_{i,j'}(\ba) : V_{i,j'} \in \bpa(V_{i,j}) \setminus \bA \}\big)$; here, $\ell$
is a generic index that sweeps over main variables in the
graph. This means that if a parent of $V_{i,j}$ is directly manipulated, it is
set to the corresponding value in $\ba$. Otherwise, the parent takes on its
potential value after intervention on any causally prior variables, or its natural value otherwise. To obtain the parent's
potential value, we follow the same definition recursively. When defining potential outcomes, intervention on $V_{i,j}$ itself is ignored. To illustrate, consider the
mediation graph of Figure~\ref{fig:mediation}(b). Possible potential outcomes
for $V_{i,3}$ are (i) $V_{i,3}(\varnothing) = V_{i,3}(V_{i,1}, V_{i,2})$, the observed distribution; (ii)
$V_{i,3}(v_{i,1}) = V_{i,3}(v_{i,1}, V_{i,2}(v_{i,1}))$, relating to total effects; and (iii)
$V_{i,3}(v_{i,1}, v_{i,2})$, relating to controlled effects.

%\vspace{-.3cm}
\subsection{Principal Stratification}
\label{sec:po_strata}

Analysts have little information about the disturbances $\bU_i$, which may take on an infinite number of values. This poses an analytic challenge, as it is difficult to reason about infinite spaces. Here, we review a result that makes the general partial identification problem tractable despite this issue: broadly, when $\bV_i$ are discrete, the model that a DAG encodes can be represented by a finite number of parameters without loss of generality, as long as the reduced space is sufficiently large. We then introduce the \textit{functional parameterization} used for this task, discuss its relationship to principal strata, and review how any marginal of the full data law can be represented in terms of these parameters.

\citet{wolfe2021cardinalities} show that there are finite state spaces for $\bU_i$ that do not restrict the NPSEM of a DAG for $\Pr(\bV_i=\bv)$, i.e.\ the model over the \emph{factual} main variables. In the following proposition, we extend this result to show that there are finite state spaces for $\bU_i$ that do not restrict the NPSEM of a DAG for the full data law---i.e., the full distribution over all factual and counterfactual versions of the main variables.

\vspace{-.2cm}
\begin{proposition} \label{prop:finite-u}
Suppose $\cG$ is a canonical DAG over discrete main variables $\bV_i$ and disturbances $\bU_i$ with infinite cardinality. Then the model over the full data law implied by $\cG$ is unchanged by assuming that the disturbances have finite cardinalities, provided those cardinalities are sufficiently large.
\end{proposition}

\vspace{-.2cm}
\noindent A proof can be found in Appendix~\ref{app:nonrestrictive}, along with details on how to obtain a lower bound on non-restrictive cardinalities for the disturbances.

Further, \cite{evans2018margins} showed that for a large class of graphs called \emph{geared} graphs, it is possible to develop a functional model that does not alter the causal model of a DAG. In the functional model of a graph, each main variable $V_i$ is associated with a disturbance $U_i$ that fully determines how $V_i$ responds to the values of its remaining parents.\footnote{Note that if any main variable $V_i$ has multiple parents in $\bU_i$, there may be multiple valid functional parameterizations, depending on which disturbance is chosen to determine which main variable. If each main variable has only a single parent in $\bU_i$, there is only a single functional parameterization.}

Proposition~\ref{prop:finite-u} enables us to develop functional models for graphs that are not geared as well. \cite{wolfe2021cardinalities} presents an algorithm for constructing a concise functional model for non-geared graphs, taking as input a disturbance cardinality that is non-restrictive of the model over factual random variables. By instead substituting the Proposition~\ref{prop:finite-u} disturbance cardinality, which may be larger and restricts neither the factual nor the  counterfactual random variables, we obtain a functional model that is likewise non-restrictive of the full data law. Intuitively, functional models are closely related to principal stratification
\citep{GreeRobi1986,Frangakis2002}. For example, consider the simple
DAG,
\begin{equation}
U_{i,1}~\rightarrow~V_{i,1}~\rightarrow~V_{i,2}~\leftarrow~U_{i,2} \label{eq:dag_simple}
\end{equation}

\noindent
in which
$V_{i,1}$ and $V_{i,2}$ are binary. This relationship is governed by the
structural equations $V_{i,1} = f_1(U_{i,1})$ and $V_{i,2} = f_2(V_{i,1},
U_{i,2})$, where the functions $f_1: \cS(U_{i,1}) \rightarrow \cS(V_{i,1})$ and $f_2:
\cS(V_{i,1}) \times \cS(U_{i,2}) \rightarrow \cS(V_{i,2})$ are deterministic and shared across
all units. Thus, the only source of randomness is in the disturbances, $\bU_i =
\langle U_{i,1}, U_{i,2} \rangle$. 

Analysts generally do not have direct information about these disturbances. For
example, $U_{i,1}$ could potentially take on any value in $(-\infty,
  \infty)$. However, Proposition~\ref{prop:finite-u} states that this
variation is irrelevant, because $V_{i,1}$ can only take on two possible values:
0 and 1. The space of $U_{i,1}$ can therefore be divided into two
\textit{canonical partitions} \citep{balke1997bounds}---those that
deterministically lead to $V_{i,1}=0$ and those that lead to $V_{i,1}=1$---and
thus there is no loss of generality in treating $U_{i,1}$ as if it were binary.\footnote{See Section 8.2 of \cite{pearl2009causality} for a related discussion.}

The situation with $V_{i,2}$ is similar but more involved. After the random
$U_{i,2}$ is realized, it induces the \emph{partially applied} response function
$V_{i,2} = f_2(V_{i,1}, U_{i,2}=u_{2}) = f_2^{(U_{i,2}=u_{2})}(V_{i,1})$, which
deterministically governs how $V_{i,2}$ counterfactually responds to
$V_{i,1}$. Regardless of how many values the disturbance can take on, this
response function must fall into one of only four possible groups, or
\textit{principal strata}, each corresponding to a possible function of the form
$f_2^{(U_{i,2}=u_{2})} : \cS(V_{i,1}) \rightarrow \cS(V_{i,2})$ \citep{angrist1996}. These
groups are (i) $V_{i,2}=1$ regardless of $V_{i,1}$, ``always takers''; (ii)
$V_{i,2}=0$ regardless of $V_{i,1}$, ``never takers''; (iii) $V_{i,2}=V_{i,1}$,
``compliers''; and (iv) $V_{i,2}=1-V_{i,1}$, ``defiers''. Thus, from the
perspective of $V_{i,2}$, any finer-grained variation in $\cS(U_{i,2})$ beyond
the canonical partitions is irrelevant. These partitions of $\bU$
are in one-to-one correspondence with principal strata, allowing causal quantities to be expressed in simple algebraic expressions; e.g., the average treatment effect (ATE) in \eqref{eq:dag_simple} is equal to the proportion of compliers minus the proportion of defiers.\footnote{To see this, note that the ATE is given by $\E[V_{i,2}(V_{i,1}=1) - V_{i,2}(V_{i,1}=0)] = \sum_{\rm strata} \E[ V_{i,2}(V_{i,1}=1) - V_{i,2}(V_{i,1}=0) \ | \ {\rm strata} ] \cdot \Pr(\text{strata}) = 0 \cdot \Pr( \text{always taker} ) + 0 \cdot \Pr( \text{never taker} ) + 1 \cdot \Pr( {\rm complier} ) - 1 \cdot \Pr( {\rm defier} )$.} By writing down all information in terms of (possibly unknown) strata sizes, we can convert causal inference problems into tractable polynomial programming problems over these variables.

The functional parameterization of this graph has four free parameters: one for the binary $U_{i,1}$ (or its reduced representation) and three for the quaternary $U_{i,2}$.\footnote{These can be thought of as the probabilities of encountering a unit of the ``control'' type with $V_{i,1}=0$ (for $U_{i,1}$) and of encountering units of the ``always-taker,'' ``never-taker,'' and ``complier'' types (for $U_{i,2}$). These parameters determine the probabilities of the remaining types (the ``treatment'' type for $U_{i,1}$ and the ``defier'' type for $U_{i,2}$), as principal strata probabilities must sum to unity.} Because the distributions of disturbances are independent in canonical DAGs by virtue of their exogeneity, only their marginal distributions need be parameterized. Each of $U_{i,1}$ and $U_{i,2}$ encode full information about how $V_{i,1}$ and $V_{i,2}$ respectively respond to their remaining parents. In other words, each setting of $\bU_i$ provides full information not only about each variable $\bV_i$, but also about each of its potential outcomes. This means that we can represent ``cross-world'' distributions such as $\Pr\big(V_{i,2}(V_{i,1}=0)=0, V_{i,2}(V_{i,1}=1)=1 \big)$---the ``complier'' proportion---in terms of parameters of the marginal distributions of $U_{i,2}$ alone. As we will see below, this fact will be useful in encoding cross-world type assumptions like monotonicity, as well as for bounding cross-world targets like the natural direct effect or the probability of causation. More generally, any marginal of the full data law may be expressed in terms of the functional parameters.

Finally, consider a more complex example, the mediation DAG of
Figure~\ref{fig:mediation}(b). The response functions for $V_{i,1}$ and $V_{i,2}$
remain unchanged. In contrast, $V_{i,3}$ is caused by $\bpa(V_{i,3}) = \langle
V_{i,1}, V_{i,2} \rangle$ via the structural equation $V_{i,3} = f_3(V_{i,1},
V_{i,2}, U_{i,23})$. Substituting in a realization of the disturbance,
$U_{i,23}=u_{i,23}$, will produce one of sixteen response functions of the form
$f_3^{(U_{i,23}=u_{23})} : \cS(V_{i,1}) \times \cS(V_{i,2}) \rightarrow \cS(V_{i,3})$. More
generally, the number of unique response functions grows with (i) the
cardinality of the variable, (ii) the number of causal parents it has, and (iii)
the parents' cardinalities. Specifically, $V_{i,j}$ has
$|\cS(V_{i,j})|^{|\cS(\bpa(V_{i,j}))|}$ possible response functions: given a
particular input from $V_{i,j}$'s parents, the number of possible outputs for
$V_{i,j}$ is $|\cS(V_{i,j})|$; the number of possible inputs from $V_{i,j}$'s
parents is $|\cS(\bpa(V_{i,j}))| = \prod_{V_{i,j'} \in \bpa(V_{i,j})}
|\cS(V_{i,j'})|$, the product of the parents' cardinalities.

In turn, this determines the minimal cardinality of $\bU$ in a reduced but
non-restrictive functional model---roughly speaking, the number of principal
strata combinations that exist, if we think of $\bU$ as principal strata. Here,
``non-restrictive'' means that the simplified model is fully expressive, or that
it can represent any possible full data law. For example, to capture the joint response patterns that a unit may
have on $V_{i,2}$ and $V_{i,3}$, a reduced version of $U_{i,23}$ will be capable of expressing any full data law if it has a cardinality of $|\cS(U_{23})| = 4 \times 16$, because $V_{i,2}$ has
four possible response functions and $V_{i,3}$ has sixteen.

\section{Formulating the Polynomial Program}\label{sec:poly}
%\label{sec:formulate_program}

We now turn to the central problem of this paper: sharply bounding causal quantities with missing data. Our approach is to (i) rewrite the causal query into a polynomial expression, (ii) rewrite modeling assumptions and empirical information into polynomial constraints, and (iii) thereby transform the task into a constrained optimization problem that can be solved computationally. \textit{Sharp bounds} are the narrowest range
that contain all admissible values for a target quantity, i.e., all values
that are consistent with information available to the analyst: structural causal
knowledge in the form of a canonical DAG, $\cG$; as well as empirical evidence,
$\cE$, and modeling assumptions, $\cA$, formalized below. We also suppose that
the main variables take on values in a known, discrete set, $\cS = \cS(\bV)$. In
this section, we will demonstrate (i) that $\{\cG, \cE, \cA, \cS \}$ restricts the admissible values of the target quantity, and (ii)
this range of observationally indistinguishable values can be recovered by polynomial
programming.

The causal graph and sample space,
$\cG$ and $\cS$, together imply a set of possible functional models, each fully characterizing the
main variables. By Proposition~\ref{prop:finite-u}, without loss of
generality, we can consider a simple functional model in which (i)
each counterfactual main variable is a deterministic function of exogeneous,
discrete disturbances; (ii) there are a relatively small number of such
disturbances; and (iii) disturbances take on a finite number of
possible values, corresponding to principal strata of the main variables. When
repeatedly sampling units (along with each unit's random disturbances, $\bU_i$),
the $k$-th disturbance thus follows the categorical distribution with parameters
$\cP_{U_k} = \langle \Pr(U_{i,k} = u_{i,k}) : u_{i,k} \rangle$. By the
properties of canonical DAGs, these disturbances are independent. It follows
that the parameters $\cP_{\bU}$ of the joint disturbance distribution
$\Pr(\bU_i=\bu_i) = \prod_k \Pr(U_{i,k}=u_{i,k})$ not only fully determine the
distribution of each factual main variable---i.e.\ the potential outcome under
no intervention, $V_{i,j}(\varnothing)$---they also determine the counterfactual
distribution of $V_{i,j}(\ba)$ under any intervention $\ba$, and its
joint distribution with other counterfactual variables $V_{i,j'}(\ba')$ under
possibly different interventions $\ba'$. This leads to the
following proposition.% about marginal probabilities of counterfactuals.

\begin{proposition} \label{prop:polynomialization}
  Suppose $\cG$ is a canonical DAG and $\cC = \{ V_{i,\ell}(\ba_{\ell}) =
  v_{\ell} \}$ is a set of counterfactual statements, indexed by $\ell$, that
  variable $V_{i,\ell}$ will take on value $v_{\ell}$ under manipulation
  $\ba_{\ell}$. Let $\cU \subset \cS(\bU)$ indicate the subset of
  disturbance realizations that lead deterministically to every statement in
  $\cC$ being satisfied. Then under the structural equation model $\cG$,
  \begin{gather} \label{eq:polynomial}
    \Pr\l( \bigwedge_{\ell} \cC_{\ell} \r)
    =\sum_{\bu \in \cU} \prod_{u_k \in \bu} \Pr(U_{i,k}=u_k),
  \end{gather}

\noindent
which is a polynomial equation in $\cP_{\bU_i}$, the parameters of
  $\Pr(\bU=\bu)$.
\end{proposition}

For example, in the mediation setting of Figure~\ref{fig:mediation}(b),
Proposition~\ref{prop:polynomialization} implies that the joint distribution of
the factual variables---$V_{i,1}(\varnothing)$, $V_{i,2}(\varnothing)$, and
$V_{i,3}(\varnothing)$---is given by
\begin{gather}
  \Pr\big( V_{i,1}(\varnothing)=v_1, V_{i,2}(\varnothing)=v_2,
V_{i,3}(\varnothing)=v_3 \big) = \sum_{\langle u_1, u_{23} \rangle \in \cU}
\Pr(U_1=u_1) \Pr(U_{23}=u_{23}),
\end{gather}
\noindent
where $\cU = \l\{ \langle u_1, u_{23} \rangle :
f_1^{(U_1=u_1)}(\varnothing) = v_1, f_2^{(U_{23}=u_{23})}(v_1) = v_2,
f_3^{(U_{23}=u_{23})}(v_1, v_2) = v_3 \r\}$ is the set of all disturbances that
are consistent with a particular $\bV_i = \langle v_1, v_2, v_3
\rangle$. Alternatively, analysts may be interested in the probability that a
randomly drawn unit $i$ has a positive controlled direct effect when fixing the
mediator to $V_{i,2}=0$. This is given by $\Pr\big( V_{i,3}(V_{i,1}=0, V_{i,2}=0)=0,
V_{i,3}(V_{i,1}=1, V_{i,2}=0)=1 \big)$ and is similarly expressed in terms of the
disturbances as $\sum_{\langle u_1, u_{23} \rangle \in \cU'} \Pr(U_{i,1}=u_1) \Pr(U_{i,23}=u_{23})$,
summing over a different subset of the disturbance space, $\cU' = \l\{ \langle
u_1, u_{23} \rangle : f_3^{(U_{i,23}=u_{23})}(V_{i,1}=1, V_{i,2}=0) = 1,
f_3^{(U_{i,23}=u_{23})}(V_{i,1}=0, V_{i,2}=0) = 0 \r\}$.

We now expand this result to include a large class of functionals of marginal
probabilities and logical statements about these functionals.

\begin{corollary} \label{cor:functional-polynomialization}
 Suppose $\cG$ is a canonical DAG. Let $\cP_{\bV}$ denote the full data law and
 $g(\cP_{\bV})$ denote a functional of $\cP_{\bV}$ involving elementary
 arithmetic operations on constants and marginal probabilities of
 $\cP_{\bV}$. Then $g(\cP_{\bV})$ can be expressed as a polynomial fraction in
 the parameters of $\cP_{\bU}$, $h(\cP_{\bU})$, by replacing each marginal
 probability with its Proposition
 \ref{prop:polynomialization} polynomialization.
\end{corollary}

We say functionals of the full data law that fulfill these properties are
\textit{polynomial-fractionalizable}, or simply \textit{polynomializable} if the
result contains no fractions. The corollary has a number of implications, which
we briefly discuss here. First, it demonstrates that a wide array of
single-world and cross-world functionals can be expressed as polynomial
fractions. These include traditional targets such as the ATE, as well as more complex targets such as the pure direct effect and the
probability of causal sufficiency. It also suggests that any non-elementary
functional of $\cP_{\bV}$ can be approximated to arbitrary precision by a
polynomial fraction, provided the functional has a convergent power series.\footnote{We note that non-elementary functionals rarely arise in practice, with the exception of target quantities on logarithmic or exponential scales. In such cases, bounds on monotonic transformations of polynomials can be straightforwardly obtained by bounding the underlying polynomial, then applying the transformation. An example of a functional that our approach cannot handle is the non-analytic $\bbone(\text{ATE is rational})$.}

Next, observe that when (i) $g(\cP_{\bV})$ is a polynomial-fractionalizable
expression; (ii) $\bstar \in \{ <, \le, =, >, \ge, \ne \}$ is a binary
comparison operator; and (iii) $\alpha$ is a constant, then statements of the
form $g(\cP_{\bV})~\bstar~\alpha$ can be equivalently expressed as
non-fractional polynomial relations $h(\cP_{\bU})~\bstar~0$. Finally, by the same token, any
polynomial-fractional expression $h(\cP_{\bU})$ in the parameters of $\cP_{\bU}$
can be reexpressed with (i) a non-fractional polynomial in an expanded parameter
space and (ii) a polynomial equation in the same expanded
space.\footnote{To see this, let $s$ be a scalar auxiliary variable and set
$h(\cP_{\bU}) = s$, which can be manipulated to obtain a non-fractional
polynomial equation, per (ii). The original expression can now be rewritten
simply as $s$, which is a monomial and hence a polynomial, per (i). Thus, the
original polynomial-fractional expression has been reexpressed in terms of (i) a
non-fractional polynomial expression and (ii) a non-fractional polynomial
equation.} We will make extensive use of these properties to convert causal queries to polynomial programs.

In Appendix~\ref{app:algorithms}, Algorithm~\ref{alg:polynomial-program-simple} provides a step-by-step procedure for obtaining sharp bounds. We begin by transforming a factual or counterfactual target of inference $\cT$ into polynomial form, possibly with the use of additional auxiliary variables to eliminate fractions. To accomplish this task, the procedure utilizes the possibly non-canonical DAG $\cG$ and the possible main-variable outcomes $\cS(\bV)$ to reexpress $\cT$ in terms of functional parameters that correspond to principal strata proportions. The result is the objective function of the polynomial program. The procedure then polynomializes the sets of constraints on polynomializable functionals resulting from empirical
evidence and by modeling assumptions, respectively  $\cE$ and $\cA$. For example, in the binary
mediation setting of Figure~\ref{fig:mediation}, $\cG$ may be the graph depicted
in either panel (a) or (b). If only observational data is available, then $\cE$
consists of eight pieces of evidence, each represented as a statement
corresponding to a cell of the factual distribution
$\Pr\big(V_{i,1}(\varnothing)=v_1, V_{i,2}(\varnothing)=v_2, V_{i,3}(\varnothing)=v_3\big) =
\Pr(V_{i,1}=v_1, V_{i,2}=v_2, V_{i,3}=v_3)$ for observable values in $\{0, 1\}^3$. Modeling
assumptions include all other information, such as monotonicity or dose-response
assumptions; these can be expressed in terms of principal strata. For example, the assumed unit-level monotonicity of the $V_1
\rightarrow V_2$ relationship \citep[e.g., the ``no defiers'' assumption
  of][]{angrist1996} can be written as the statement that $\Pr\big(V_{i,2}(V_{i,1} = 0)
= 1, V_{i,2}(V_{i,1} = 1) = 0 \big) = 0$. Assumed population-level monotonicity is
typically written $\E[V_{i,2}(V_{i,1} = 1) - V_{i,2}(V_{i,1} = 0)] \ge 0$, but can equivalently
be reformulated in terms of principal strata as $\Pr\big(V_{i,2}(V_{i,1} = 1) = 1, V_{i,2}(V_{i,1} = 0) = 0 \big) -
\Pr\big(V_{i,2}(V_{i,1} = 0) = 1, V_{i,2}(V_{i,1} = 1) = 0 \big) \ge 0$. Finally, the statement that each disturbance $k$ follows a categorical probability
distribution is reexpressed as the polynomial relations $\Pr(U_k=u_k) \ge 0 : u_k$ and $\sum_{u_k} \Pr(U_k=u_k) = 1$.

Algorithm~\ref{alg:polynomial-program-simple}
produces an optimization problem with a polynomial objective subject to
polynomial constraints. This polynomial programming problem is equivalent to the
original causal bounding problem. This leads directly to the following theorem.

\begin{theorem}
  \label{thm:sharp}
 Minimization (maximization) of the polynomial program produced by Algorithm
 \ref{alg:polynomial-program-simple} produces sharp lower (upper) bounds on
 $\cT$ under the sample space $\cS(\bV)$, structural equation model $\cG$,
 additional modeling assumptions $\cA$, and empirical evidence $\cE$.
\end{theorem}

Once the causal problem is expressed in polynomial form, a variety of computational solvers can in principle be used to optimize \citep[e.g.\ IPOPT;][]{wachter2006implementation}. However, local solvers cannot guarantee valid bounds without exhaustively searching the space; when time is limited, these often fail to discover global extrema for the causal estimand, resulting in intervals that may fail to contain the quantity of interest. Moreover, such approaches often become computationally intractable as causal problems grow complex. In the next section, we show how the polynomial program can be simplified to speed computation.

\section{Simplifying the Polynomial Program}\label{sec:simp}
%\label{sec:simplification}

Because solving polynomial programs is in general NP-hard, efficient computation
requires us to fully exploit our knowledge of the problem structure. This
knowledge allows analysts to reduce the complexity of the program in ways that
algebraic presolvers may not necessarily detect. In this section, we discuss
several ways to do this.

\subsection{Eliminating Blocked Disturbances}

We begin by using the following observation to limit the number of disturbance
distribution parameters involved in the target and constraints.

\begin{proposition} \label{prop:relevant-latents}
  Consider the polynomialization of a probability $\Pr\Big(\bigwedge_{\ell} \cC_{\ell}
  \Big)$, where $\cC = \{ V_{i,\ell}(\ba_{\ell})=v_{\ell} : \ell \}$. We say that for
  intervention $\ba_{\ell}$, a disturbance $U_{i,k}$ is blocked from the
  corresponding counterfactual $V_{i, \ell}$ if there are no paths from
  $U_{i,k}$ to $V_{i, \ell}$ that do not go through the causally prior members
  of the intervention $\ba_{\ell}$. When $U_{i,k}$ is blocked from
  $V_{i,\ell}$ for every $\ell$, the corresponding parameters
  $\cP_{U_k}$ can be eliminated from the polynomialization.
\end{proposition}

In other words, Proposition~\ref{prop:relevant-latents} states that each main
variable $V_{i,j}$ is only a function of its ancestors in $\bU$ that affect it
through a variable not under intervention. For each marginal probability of an
event, the disturbances that do not affect any variable in the event are
irrelevant. This allows us to amend the polynomialization of Proposition
\ref{prop:polynomialization} so that the outer sum ranges only over all possible
settings of \textit{relevant} disturbances, reducing the degree of each term
in the polynomial. For example, in the mediation graph of
Figure~\ref{fig:mediation}(b), consider the total effect of the treatment
$V_{i,1}$ on the outcome $V_{i,3}$. Here, all probabilities are of the form
$V_{i,3}(v_{i,1}=a_1)=v_3$.\footnote{The total effect is given by
$\Pr\big( V_{i,3}(V_{i,1}=1)=1 \big) - \Pr\big( V_{i,3}(V_{i,1}=0)=1 \big)$, which can
equivalently be written $\Pr\big( V_{i,3}(V_{i,1}=1, V_{i,2}=V_{i,2}(V_{i,1}=1))=1 \big) -
\Pr\big( V_{i,3}(V_{i,1}=0, V_{i,2}=V_{i,2}(V_{i,1}=0))=1 \big)$.} The disturbance $U_{i,1}$ is
therefore blocked from the outcome $V_{i,3}$, because the sole path from
$U_{i,1}$ to $V_{i,3}$ goes through the intervention set $V_{i,1}$. This means
that whenever $\Pr(U_1=u_1)$ appears in the polynomial, it does so in a way that
ensures $\sum_{u_1} \Pr(U_1=u_1)=1$ can be factored out and eliminated.

% Next, we provide two less straight-forward methods for simplifying the program.

\subsection{Exploiting the Nested Markov Parameterization}

We now consider the common case when the empirical evidence $\cE$ includes
\textit{single-world marginal distributions}. This occurs when (i) a factual or
counterfactual event $\bigwedge_\ell \{ V_{i,\ell}(\ba)=v_\ell \}$ involves the
same intervention, $\ba$, for every variable of interest; and (ii) the
probability $\Pr\Big( \bigwedge_\ell \{ V_{i,\ell}(\ba)=v_\ell \} \Big)$ is
observed for every event in that state space, $\langle v_\ell : \ell \rangle \in
\prod_\ell \cS(V_{i,\ell})$. For example, in the binary mediation setting of
Figure~\ref{fig:mediation}(b), an observational study might obtain information
about $\Pr\big( V_{i,1}(\varnothing)=v_1, V_{i,2}(\varnothing)=v_2,
V_{i,3}(\varnothing)=v_3 \big)$ for every combination of $v_1$, $v_2$, and
$v_3$. Similarly, an experiment that randomly manipulated
$V_{i,1}$ would obtain two such distributions: (i) by observing $\Pr\big(
V_{i,2}(V_{i,1}=0)=v_2, V_{i,3}(V_{i,1}=0)=v_3 \big)$ for all $v_2$ and $v_3$;
and similarly, (ii) by observing $\Pr\big( V_{i,2}(V_{i,1}=1)=v_2,
V_{i,3}(V_{i,1}=1)=v_3 \big)$ for all $v_2$ and $v_3$.

A na\"ive parameterization might use one parameter for the probability of each principal strata (e.g., four parameters for the proportion of ``always takers,'' ``never takers,'' ``compliers,'' and ``defiers''). 
It is immediately apparent that one na\"ive parameter is redundant, as its value is already implied by the fact that all distributions must
marginalize to unity. Hidden-variable DAG models imply additional equality constraints, each of which can likewise be used to reduce the number of parameters needed to describe the model, thereby reducing the number of polynomial constraints in the program.

These additional equality constraints, which are implied by the structural
equations model of $\cG$, are well understood. In particular, $\cG$ imposes
certain \textit{conditional independence} and \textit{generalized equality}
  constraints \citep[or Verma
  constraints,][]{verma1990_equivalence,tian2002_testable} on these
distributions. Each equality constraint can be used to eliminate one parameter of the single-world distribution. The parameterization that takes full
advantage of these structural equality constraints to reduce the number of
parameters is called the \textit{nested Markov} parameterization \citep{evans2019smooth}. This parameterization achieves the minimal number of
parameters, equal to the dimension of the model of $\cG$.

Each nested Markov parameter is exactly equal to an identified marginal probability of a
single-world event, $\Pr\Big( \bigwedge_{\ell'} \{ V_{i,\ell'}(\ba')=v_{\ell'}
\} \Big)$. Because each of the nested
Markov parameters is identified from the initial single-world distribution,
$\Pr\Big( \bigwedge_\ell \{ V_{i,\ell}(\ba)=v_\ell \} \Big)$, whether indirectly
or directly, it can be calculated directly from the empirical evidence $\cE$. By Proposition \ref{prop:polynomialization}, this probability
remains polynomializable in the parameters $\cP_{\bU}$. This allows us to add
one equality constraint to the program per nested Markov parameter, 
based on its polynomialization, rather than one equality constraint per outcome in the state space. For hidden variable DAGs that imply a large number of equality constraints, this can substantially reduce the number of constraints. \cite{evans2019smooth} offers a a complete guide to obtaining nested Markov formulations of arbitrary single-world distributions.

Each parameter in the nested Markov formulation is the probability of a single-world
event that involves fewer main variables and more interventions, compared to any event used by the na\"ive parameterization. As a result, the corresponding polynomialization will have fewer terms, of lower degree, than the polynomialization of na\"ive parameters. An example is provided in
Appendix~\ref{app:nested_markov_example}. Using this technique, we modify
Algorithm \ref{alg:polynomial-program-simple} by partitioning the empirical
evidence into all single-world marginal distributions $\cE_M$ and the remaining
evidence $\cE_R$. The constraints in $\cE_M$ can then be reduced into their
nested Markov form before polynomialization. Because the nested Markov parameterization allows for fewer, simpler polynomial constraints in the program, it is important to use it whenever the empirical evidence permits.

We note that when certain deterministic relationships
exist between variables in $\bV_i$, as in the missing-data setting of
Figure~\ref{fig:common}(c--d),\footnote{In this graph, a latent
variable $Y$ has an observed version $Y^\ast$ that deterministically inherits
$Y^\ast=Y$ when a reporting variable $R=1$, but takes on the
missing-value indicator $Y^\ast=\texttt{NA}$ otherwise.} these relationships may
imply equality constraints not exploited by the nested Markov
parameterization. In such cases, it may be possible to further reduce the number of constraints; we do not explore that option here.

\subsection{Eliminating Additional Constraints and Parameters}

Finally, we describe when constraints and parameters can be safely eliminated
from a program. We say that parameters $x$ and $y$ \textit{co-occur} in a polynomial system if they appear in the same
constraint; they \textit{interact} if there exists a sequence of parameters from $x$ to $y$ such
that every adjacent pair  co-occurs.\footnote{For example, consider the constraints $x + y = a$,  $y + z = b$. Here, $x$ and $y$ co-occur; $x$ and $z$ interact.} If a constraint's parameters do not interact with the objective's parameters, that constraint may be dropped. If a
parameter exists only in constraints that have been eliminated, then the parameter has also been eliminated, simplifying the system.

This may be used in conjunction with the structure of $\cG$ to help
simplify the program, because different \textit{districts}---components in $\cG$ connected by bidirected arcs \citep{tian2002_testable,richardson2003markov}---do typically do not interact. That is, likelihoods on marginal distributions of $\cG$ have a representation that is decomposable by districts. For example, in Figure~\ref{fig:mediation}(b),
$V_{i,1}$ lies in one district; in contrast, $V_{i,2}$ and $V_{i,3}$ lie in
another district, because they are connected by $U_{i,23}$. Because each nested Markov parameter is the probability of a single-world
event involving main variables within a
single district, its polynomialization will at most involve disturbance parameters in that district. This leads to the following proposition.

\begin{proposition} \label{prop:constraint-complexity}
The degree of each polynomial in the nested Markov constraints is bounded from
above by the number of latent variables in the corresponding district. Moreover,
if two disturbances $U_{i,k}$ and $U_{i,k'}$ appear in different districts, their
parameters $\cP_{U_k}$ and $\cP_{U_{k'}}$ will not interact in any nested Markov
constraint.
\end{proposition}

To illustrate, consider the common scenario
where an analyst observes the full joint distribution over factual variables,
$\Pr\big( V_{i,1}(\varnothing)=v_1, \ldots, V_{i,J}(\varnothing)=v_J \big)$, and
seeks to bound a functional relating a treatment $\ba$ to an outcome
$V_{i,j}(\ba)$ in the same district. As an example, in
Figure~\ref{fig:mediation}(b), the effect of the mediator $V_{i,2}$ on the
outcome $V_{i,3}$ is wholly contained within a single district. We
can therefore drop all constraints related to nested Markov parameters involving other
districts, and thus all disturbance parameters in other districts.

\section{Computing $\varepsilon$-sharp Bounds in Polynomial Programs}
\label{sec:algo}

We now turn to the practical optimization of the polynomial program defined by
Algorithm~\ref{alg:polynomial-program-simple}. Theorem~\ref{thm:sharp} states
minimization and maximization of this original primal program is equivalent to the initial
bounding problem. However, obtaining globally optimal
solutions in polynomial programming can be computationally intensive. Worryingly, methods
that iteratively improve suboptimal values for the primal problem may fail to produce valid bounds (i.e., bounds containing all possible values of the estimand, including global
extrema) without searching the full parameter
space, $\cP$. To address
this challenge, we use \textit{dual} methods that construct and iteratively
refine an outer envelope around the primal function (i.e.\ the objective function, or causal quantity of interest). Specifically, we employ a
variation of the spatial branch-and-bound method, combined with a piecewise
linear envelope, implemented using a variety of optimization frameworks that include SCIP and Couenne
\citep{vigerske2018scip,gamrath2020scip, belelimawa08}. Throughout the optimization process, current suboptimal values for dual
minimization and maximization problems are guaranteed to produce valid but loose \textit{outer} bounds; current suboptimal values for the primal problem produce possibly invalid \textit{inner} bounds; and the lower (upper) endpoint of the unknown sharp bounds is guaranteed to lie between the current suboptimal primal and dual minimization (maximization) values. Through
simultaneous \textit{primal-dual} optimization, we use these suboptimal inner bounds to precisely quantify worst-case looseness, $\varepsilon$, of the suboptimal but
valid outer bounds. This allows researchers to assess how more computation may lead to tightened conclusions.

A step-by-step description of our optimization procedure, which we term $\varepsilon$-sharp bounding, is
given in Algorithm~\ref{alg:esharp_bound} of Appendix~\ref{app:algorithms}. At a high level, it proceeds as follows. Our procedure takes as inputs the
polynomialized objective function $\cT(\bp)$ and constraint set $\cC(\bp)$,
obtained from Algorithm~\ref{alg:polynomial-program-simple}. It then evaluates a range of models, or points $\bp$ in the model space $\cP$ for which $\cC(\bp)$ is satisfied. It seeks to identify extreme values of $\cT(\bp)$ within this subspace. It also
accepts two parameters: $\epsilon^{\text{thresh}}$, a stopping threshold for the
looseness factor stopping, and $\theta^{\text{thresh}}$, a stopping threshold
for width of the bounds. The algorithm returns two types of information: the bounds for the causal program, and the worst-case looseness
factor $\varepsilon$.

Primal bounds are denoted $\underline{P}$ and $\overline{P}$, adopting the
convention that underlines refer to objects used for minimization and overlines
for maximization. These indicate the extreme values of the target estimand in
any admissible model---that is, satisfying $\cC(\bp)$---that has been located so
far. These are initialized at $+\infty$ and $-\infty$, respectively, indicating
that no admissible models have been found yet. As optimization proceeds, the
primal bounds improve as new, more extreme admissible models are found. We refer
to $[\underline{P}, \overline{P}]$ as the \textit{inner bounds}: the unknown
sharp bounds must at least contain these points, which correspond to models that
are observationally indistinguishable from the true DGP.

Dual optimization begins by partitioning the parameter space into
branches, proceeding separately for the lower and upper bound and
respectively producing partitions $\underline{\cB}_b$ and $\overline{\cB}_b$. At
initialization, these consist of a single branch spanning the entire
parameter space; each branch is then recursively divided. The lower and upper
parts of the dual envelope, or outer envelope, are denoted $\underline{\cD}$ and
$\overline{\cD}$. These are piecewise linear functions, with pieces
corresponding to the branching partitions, that are \textit{relaxations} of the
true objective function, $\cT(\bp)$, from below and above.  These relaxations are made to ensure they will always
contain the entire objective function at all points in the parameter
space. Within branch $b$,
the value $\min \{ \underline{D}_b(\bp) : \bp \in \overline{\cB}_b \}$ indicates
the lowest value attained by the lower envelope; thus, $\underline{\cT} = \min_b
\l\{ \min \{ \underline{D}_b(\bp) : \bp \in \overline{\cB}_b \} \r\}$ represents
the lowest value attained by the lower envelope anywhere in the parameter
space. Conversely, $\overline{\cT} = \max_b \l\{ \max \{ \overline{D}_b(\bp) :
\bp \in \overline{\cB}_b \} \r\}$ represents the highest value of the upper
envelope. These extreme points on
the dual envelope, $[\underline{\cT}, \overline{\cT}]$, define the dual (outer) bounds. These are the
reported causal bounds; whatever the true
sharp bounds, they must lie inside the dual
bounds, even if the algorithm has not run to completion. We let $\theta$ equal the bound width, or the
difference between the upper and lower dual bounds, and we define the worst-case
looseness factor $\varepsilon$ as the slack (the difference in dual and
primal bound widths) divided by the primal bound width.

The algorithm heuristically selects branches in the model space that appear
promising, and refines primal and dual bounds in turn. It first
searches within the branch for an admissible model; if found, and if the associated causal estimand
is more extreme than those previously encountered, it is stored as a new primal bound. Whatever the true
nonparametric sharp bounds, they must lie outside the primal
bounds because the true bounds must contain the extreme models that define the primal bounds. Then, it
divides the branch into sub-branches and refines the dual envelope by tightening
the piecewise linear outer-approximation. The algorithm
continuously prunes branches of $\underline{\cB}_b$ and $\overline{\cB}_b$ that
wholly violate constraints; it also continuously
branches and refines the bounds while $\theta$ and $\varepsilon$ exceed
specified thresholds.

\section{Statistical Inference}\label{sec:inference}

We now turn to statistical inference for the bounds developed above. We say that
the results of Algorithm~\ref{alg:esharp_bound} when applied to $\cE$, the population empirical constraints---i.e., margins of the full data law that are observed without sampling error---are
\textit{population bounds}. In practice, the empirical quantities used in these constraints are estimated from finite samples. Our goal in this section is to account for variation in $\hat{\cE}$,
the estimated constraints, that arises over repeated
sampling. The results of Algorithm~\ref{alg:esharp_bound} when substituting
$\hat{\cE}$ for $\cE$ are referred to as the \textit{estimated bounds}. In this
section, we describe how to construct \textit{confidence bounds} that (i)
contain the estimated bounds and (ii) contain the population bounds at a rate of at least the confidence level $\alpha$ over repeated samples.

Recall that each element of empirical evidence $\cE$ is a relation between (i) some population
quantity that is an observable functional of the main variables' distribution,
$g(\cP_{\bV})$, reexpressed in terms of the disturbance distribution
$\cP_{\bU}$; and (ii) the population value of that observable quantity. In
$\hat{\cE}$, we plug in for (ii) the estimated value of the quantity in
finite data. For example, in the mediation graph of
Figure~\ref{fig:mediation}(b), an analyst with access to a sample of
observational data would have

{
  \singlespacing
\begin{gather}
\hat{\cE} =
\l\{ {\rm polynomialize}\l(
\begin{array}{l}
  \Pr\big( V_{i,1}(\varnothing)=v_1,
  V_{i,2}(\varnothing)=v_2, V_{i,3}(\varnothing)=v_3 \big) \\
  \quad = \frac{1}{N}
  \sum_{i=1}^N \bbone\l\{ V_{i,1}=v_1, V_{i,2}=v_2, V_{i,3}=v_3 \r\}
\end{array}
\r) : v_1, v_2, v_3 \r\}
\end{gather}
}

\noindent We will refer to the vector of estimated quantities on the right-hand side of
$\hat{\cE}$ elements---in the above example, quantities of the form $\frac{1}{N}
  \sum_{i=1}^N \bbone\l( V_{i,1}=v_1, V_{i,2}=v_2, V_{i,3}=v_3\r)$---as $\hat{\bE}$. We denote the corresponding population quantities
as $\bE$, the right-hand side values in $\cE$.

To construct confidence bounds we consider the sampling variability of these estimated quantities. We construct
regions, ${\rm CR}_\alpha(\hat{\bE})$, containing $\hat{\bE}$ and guaranteed to contain the population quantities $\bE$ with
at least probability $\alpha$ over repeated samples. These regions correspond to population distributions over observed parameters that cannot be rejected at level $\alpha$. In Algorithm~\ref{alg:esharp_bound}, we then replace the
$\hat{\cE}$ constraints with a set of loosened \textit{confidence constraints}
${\rm CR}_\alpha(\hat{\cE})$. In other words, if the population bounds are obtained by optimizing subject to a equality constraint $\l\{ g_\ell(\cP_{\bV}) = \bE_\ell \r\} \in \cE$, and the estimated bounds are obtained with the plug-in version $\l\{ g_\ell(\cP_{\bV}) = \hat{\bE}_\ell \r\} \in \hat{\cE}$, then the confidence bounds will incorporate the interval constraint $\l\{ g_\ell(\cP_{\bV}) \in {\rm
  CR}_\alpha(\hat{\bE}_\ell) \r\} \in \rm{CR}_\alpha(\hat\cE)$. 
  
Because loosening $\hat\cE$ to $\rm{CR}_\alpha(\hat\cE)$ can only decrease (increase)
the minimum (maximum) value obtained by the polynomial program,
confidence bounds always contain the estimated bounds. Similarly, when
the confidence region for estimated quantities fully contains their population analogues, then the confidence constraint $\rm{CR}_\alpha(\hat\cE)$ is looser than the population constraint $\cE$, and resulting confidence bounds also contain the population bounds. However, when the confidence region does not fully contain population quantities due to sampling error, confidence bounds may still contain population bounds. This can occur if the non-covered quantity corresponds to a constraint that is irrelevant to the bounds. Therefore, if the confidence region on the observed quantities has coverage of exactly $\alpha$, confidence bounds will contain the population bounds in at least $\alpha$ of repeated samples.

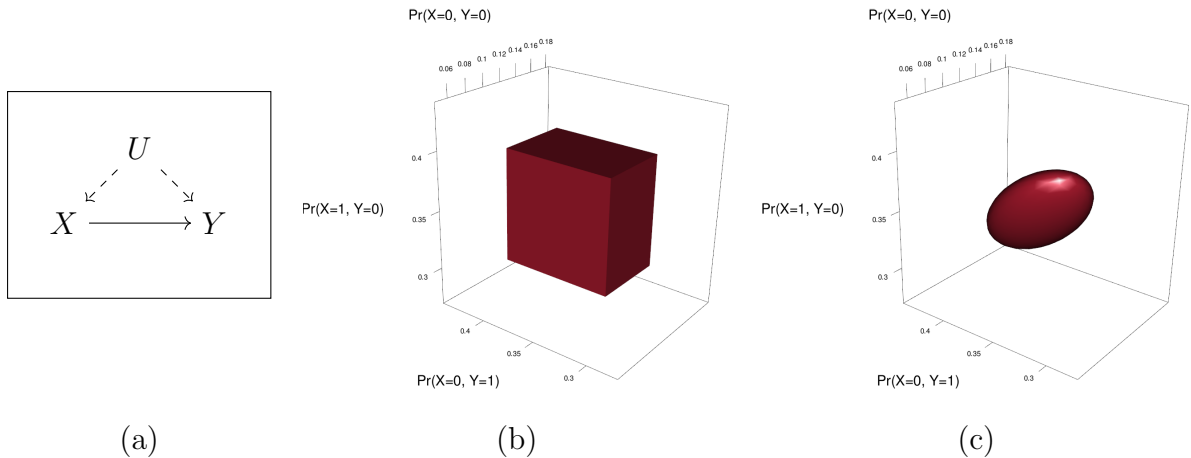
\begin{figure}
    \caption{ \textbf{Polynomial confidence regions in a binary graph.} Panel (a) presents a causal graph in which binary $X$ causes binary $Y$, but both are confounded by an unobserved $U$. $N=1,000$ observations are drawn from this DGP, producing an empirical distribution with proportions $\frac{1}{N} \sum_{i=1}^N \bbone(X_i=x, Y_i=y)$. Panels (b--c) depict confidence regions for $\Pr(X_i=0, Y_i=0)$, $\Pr(X_i=0, Y_i=1)$, and $\Pr(X_i=1, Y_i=0)$; the final category, $\Pr(X_i=1, Y_i=1)$ (not depicted), must sum to unity. Panel (b) shows the Bernoulli-KL confidence region, which is conservative in finite samples and can be polynomialized as a set of linear inequalities. Panel (c) shows the Gaussian confidence region, which is asymptotically valid and can be polynomialized as a single convex quadratic inequality.}
  \label{fig:uncertainty}
  \centering
%  \vspace{2ex}
  \begin{tabular}{ccc}
 \begin{tikzpicture}%[node distance=.5 cm and .5 cm]
 \node (U) at (3,1) {$U$};
 \node (X) at (2,0) {$X$};
 \node (Y) at (4,0) {$Y$};
 \path[dashed] (U) edge [->](X);
 \path[dashed] (U) edge [->](Y);
 \path (X) edge [->](Y);
\draw (1.25,-1) rectangle (4.75,1.75);
\useasboundingbox (1.25, -2.5) rectangle (4.75,1.75);
 \end{tikzpicture}
    &
    \includegraphics[width=.35\textwidth, trim={300 50 585 50}, clip]{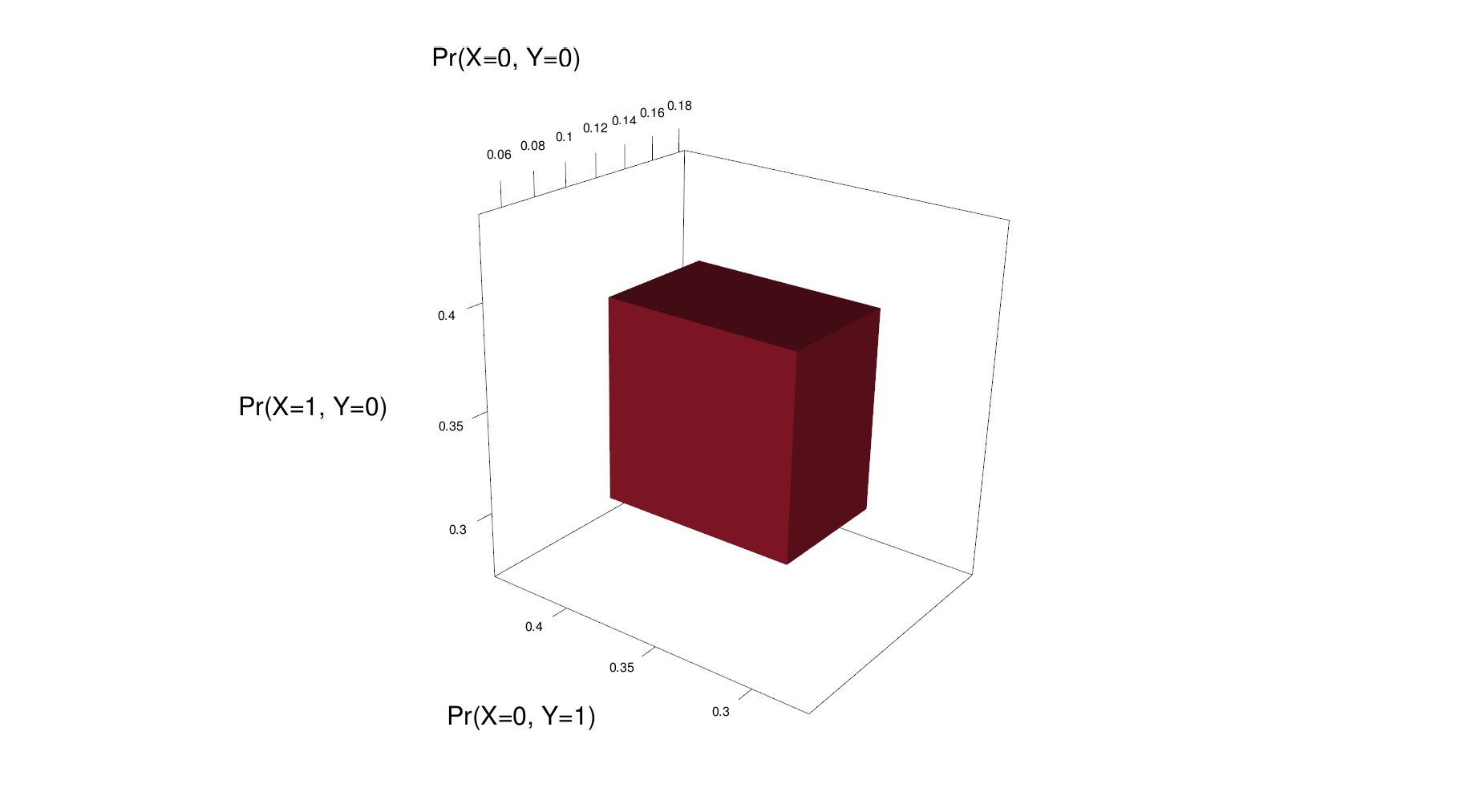}
    &
    \includegraphics[width=.35\textwidth, trim={300 50 585 50}, clip]{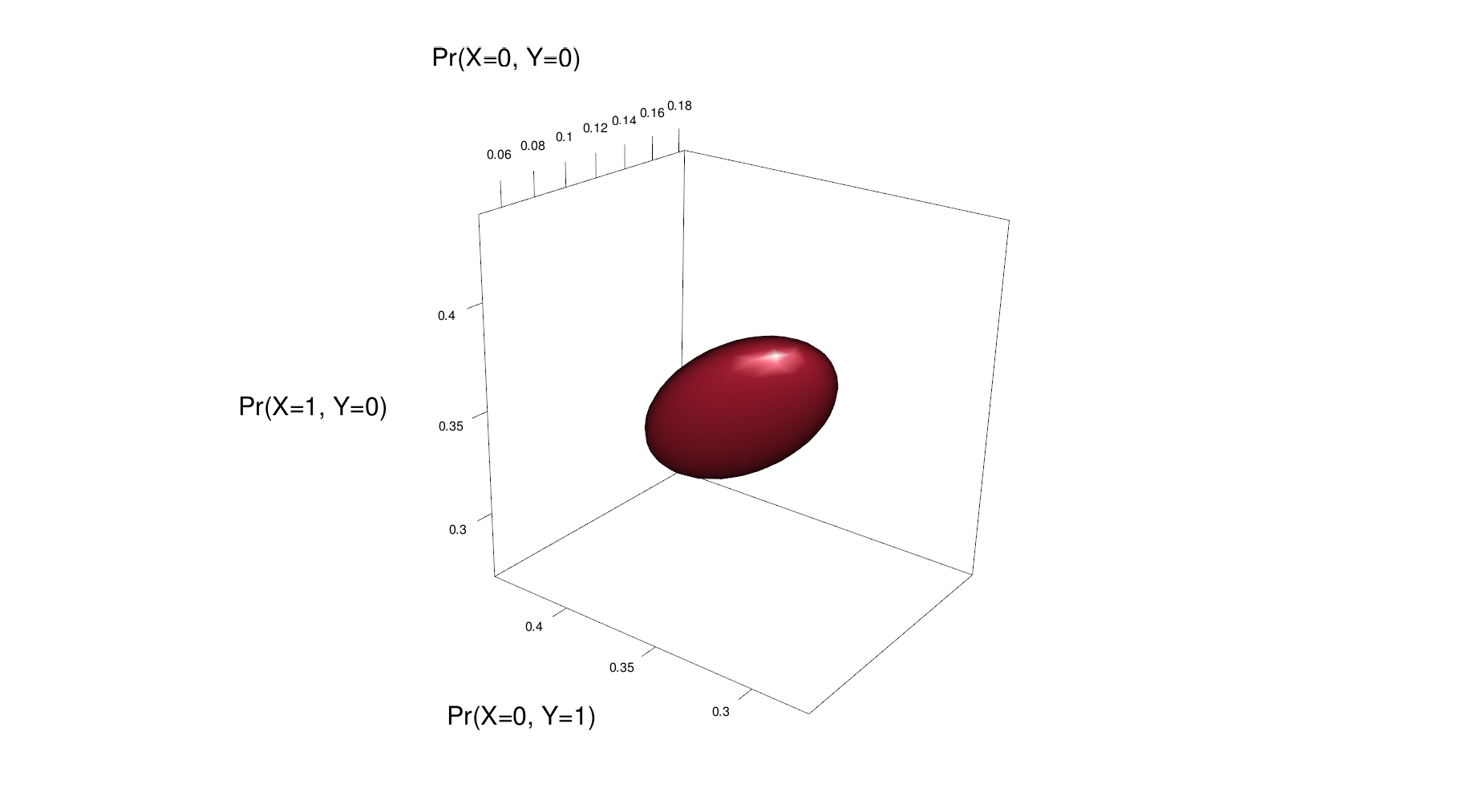}
    \\
    (a)
    &
    (b)
    &
    (c)
  \end{tabular}
\end{figure}

In discrete settings, the task of obtaining confidence bounds thus reduces to the problem of constructing regions ${\rm
  CR}_\alpha(\hat{\bE})$ for the multinomial proportion, such that $\Pr\l( \bE \in {\rm CR}_\alpha(\hat{\bE})
\r)$ $\ge \alpha$. We focus on two methods for doing so. Drawing on \citet{malloy2020optimal}, we first consider a ``Bernoulli-KL'' approach that constructs separate confidence regions for each observable atomic event, $\Pr(\bV_i = \bv)$, treating it as a ``success'' in a Bernoulli distribution. The approach rotates through all possible $\bv$ and combines the event-specific regions using a result on the Kullback-Leibler divergence of sampling distributions to the underlying population distribution. The Bernoulli-KL method produces a confidence region for single-world distributions that is guaranteed to have conservative coverage for the multinomial proportion in finite samples. The region can be represented as a system of linear inequality constraints, then incorporated into the polynomial program. Our second approach uses an asymptotically valid confidence region based on the multivariate Gaussian limiting distribution of the Dirichlet \citep{bienayme1838memoire}, which can be represented as a single convex quadratic inequality constraint. Figure~\ref{fig:uncertainty} visualizes these regions for a simple two-node graph. 
 Simulations reported in Section~\ref{sub:coverage} evaluate coverage of the methods for various sample sizes. Appendix~\ref{app:uncertainty} provides details on the implementation of these methods. We also provide a method for polynomializing arbitrary confidence regions, allowing analysts to exploit tighter finite-sample confidence regions.

\section{Simulated Examples}\label{sec:sims}
%\label{sec:simulated}

We now demonstrate our algorithm's performance via simulations. Several examples correspond to known analytic solutions, offering further validation of our approach. Section~\ref{sec:iv_sim} illustrates how Algorithms~\ref{alg:polynomial-program-simple}--\ref{alg:esharp_bound} allow analysts to iteratively state possible assumptions, test their observable implications, and use them to narrow causal bounds under noncompliance. Section~\ref{sub:coverage} evaluates our proposals for statistical inference with estimated bounds. Section~\ref{sub:gen_sim} examines several challenges---selection, mismeasurement, and missingness---that pose more complex threats to statistical inference. For clarity of exposition, all simulations use binary variables; our method adapts automatically to categorical variables.

\subsection{Instrumental Variables}\label{sec:iv_sim}

Noncompliance, or deviation between assigned ($Z_i$) and realized ($X_i$) treatment status, is a common obstacle to causal inference in randomized trials. \cite{balke1997bounds} showed that the task of bounding the ATE on an outcome $Y_i$ in the presence of noncompliance can be formulated as a linear programming problem, admitting a computational solution to partial identification. However, this approach cannot be extended to bound the local average treatment effect (LATE) among ``compliers'' that accept the assigned treatment---a principal effect that has received considerable attention---because this estimand corresponds to a nonlinear objective function. \cite{angrist1996} shows the LATE can be point identified, but only if a number of conditions hold. These conditions include (i) ignorability of $Z_i$; (ii) a non-null effect of $Z_i$ on $X_i$; (iii) an exclusion restriction, or the absence of a direct effect of $Z_i$ on $Y_i$; and (iv) monotonicity, or the absence of ``defiers'' that behave inversely to instructions. In this section, we estimate both the ATE and LATE in settings where assumptions i--ii are satisfied, then probe the implications of assumptions iii--iv.
%The Stable Unit Treatment Value Assumption (SUTVA) is also assumed throughout. ``Compliers'' are units which accept (reject) treatment as assigned. ``Defiers'' are units that accept treatment when the instrument is not assigned and reject treatment when the instrument is assigned. 
Our results show that while extant methods offer solutions for specific scenarios and estimands, even minor deviations from ideal conditions can render them inapplicable or inaccurate. Below, we show how our algorithm easily accommodates these variations and complications.

%that the traditional instrumental variables estimator recovers the complier average treatment effect (CATE) under four assumptions: 

%In many applied settings, analysts may suspect one or more of these assumptions is violated and seek to bound either the CATE, or the average treatment effect (ATE). In this section, we use iterations of this familiar scenario to demonstrate how our algorithm allows users full flexibility to change estimands and/or remove or invoke assumptions as they see fit. We also show the mistaken inferences that can result when untenable assumptions are embraced. 

%analysts may have reason to believe one or more of these assumptions is violated. Here, we simulate data in which several of these assumptions are violated, and show how incorrectly invoking these assumptions could lead analysts astray. We also show the ability of our algorithm to alert analysts to extreme scenarios in which the causal query is altogether infeasible given the observed data, a feature traditional approaches would not reveal.

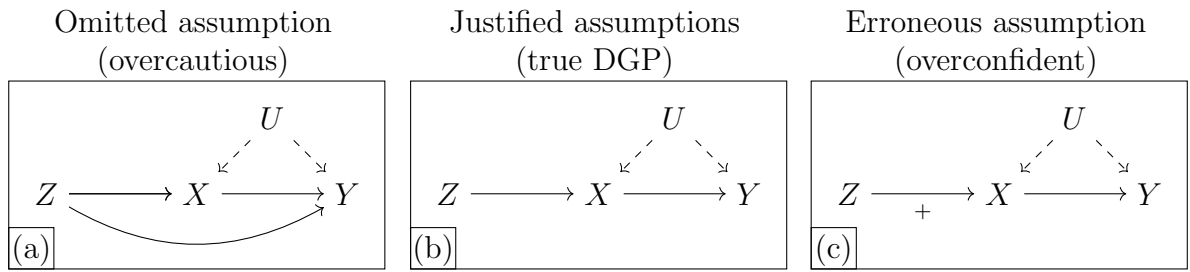
\begin{figure}[htb]
 \centering
  \caption{\textbf{DGPs with noncompliance.} The figure displays three possible causal models corresponding to scenarios in which an encouragement $Z$ causes treatment $X$. Panel (b) corresponds to the true DAG in our simulated dataset, in which the monotonicity assumption is violated (indicated here by the absence of a $+$ symbol) but other key identifying assumptions are satisfied. Panel (a) depicts a DAG assumed by an overcautious analyst that allows for violations of the exclusion restriction. Panel (c) depicts a model assumed by an overconfident analyst in which monotonicity of $Z \rightarrow X$ is incorrectly invoked.}
 \label{fig:iv}
%\vspace{3ex}
\begin{minipage}[b]{.32\textwidth}
 \centering
Omitted assumption\\
(overcautious)\\
 \begin{tikzpicture}%[node distance=.5 cm and .5 cm]
 \node (U) at (3,1) {$U$};
 \node (Z) at (0,0) {$Z$};
 \node (X) at (2,0) {$X$};
 \node (Y) at (4,0) {$Y$};
 \path[dashed] (U) edge [->](X);
 \path[dashed] (U) edge [->](Y);
 \path (Z) edge [->](X);
 \path (X) edge [->](Y);
 \path (Z) edge [->](X);
  \path (Z) edge [->, bend right=30](Y);
\node (a) at (-.2, -.75) {(a)};
\draw (-.5,-1) rectangle (.1, -.45);
\draw (-.5,-1) rectangle (4.5,1.5);
 \end{tikzpicture}
 \end{minipage}
\begin{minipage}[b]{0.32\textwidth}
 \centering
Justified assumptions\\
(true DGP)\\
 \begin{tikzpicture}%[node distance=.5 cm and .5 cm]
 \node (U) at (3,1) {$U$} ;
 \node (Z) at (0,0) {$Z$};
 \node (X) at (2,0) {$X$};
 \node (Y) at (4,0) {$Y$};
 \path[dashed] (U) edge [->](X);
 \path[dashed] (U) edge [->](Y);
 \path (Z) edge [->](X);
 \path (X) edge [->](Y);
 \node (b) at (-.2, -.75) {(b)};
\draw (-.5,-1) rectangle (.1, -.45);
\draw (-.5,-1) rectangle (4.5,1.5);
 \end{tikzpicture}
 \end{minipage}
    \begin{minipage}[b]{0.32\textwidth}
 \centering
Erroneous assumption \\
(overconfident)\\
 \begin{tikzpicture}%[node distance=.5 cm and .5 cm]
 \node (U) at (3,1) {$U$};
 \node (Z) at (0,0) {$Z$};
 \node (X) at (2,0) {$X$};
 \node (Y) at (4,0) {$Y$};
 \path[dashed] (U) edge [->](X);
 \path[dashed] (U) edge [->] (Y);
 \path (Z) edge [->] node [midway, below] {\scriptsize $+$} (X);
 \path (X) edge [->](Y);
 \node (c) at (-.2, -.75) {(c)};
\draw (-.5,-1) rectangle (.1, -.45);
\draw (-.5,-1) rectangle (4.5,1.5);
 \end{tikzpicture}
 \end{minipage}
\end{figure}

Figure \ref{fig:iv} displays three possible DGPs that analysts might assume in a scenario involving noncompliance. We simulate data from the true DGP, shown in panel (b), in which all assumptions in \cite{angrist1996} are satisfied except monotonicity of $Z \to X$. In this simulation, the true values of the ATE and LATE are $-$0.25 and $-$0.36, respectively. In practice, analysts may proceed with an abundance of caution and make the conservative causal assumptions depicted in panel (a)---a challenging scenario in which a direct effect of the instrument on the outcome cannot be excluded and monotonicity is not assumed. Assuming model (a) and applying our algorithm yields sharp bounds of $[-0.63, 0.37]$ and $[-1, 1]$ for the ATE and LATE, respectively. While these bounds are relatively wide---the ATE cannot be signed, and the bounds for LATE are entirely uninformative---the resulting intervals do contain the true estimand values, and they represent the most precise statement possible under assumptions the analyst is willing to defend.

%In addition, given the frequency of debate over whether the exclusion restriction holds in these types of applied settings, our approach offers an avenue for researchers 

If the analyst was willing to assume the exclusion restriction, per model (b)---perhaps due to domain expertise or an experimental design that ruled out direct effects---our algorithm would bound the ATE at $[-0.55, -0.15]$, revealing a negative effect and correctly containing the true value of $-$0.25. However, under these circumstances, the bounds on the LATE remain entirely uninformative at $[-1,1]$. This reflects the fact that without strong assumptions, it is difficult to learn about cross-world quantities such as principal effects.

%CATE: a) [-1,1]; b) [-1,1]; c) [-1,1]; d) infeasible ATE: a) [-0.627, +0.373]; b) [-0.627, 0.373]; c) [-0.55, -0.145]; d) infeasible

Finally, panel (c) shows a DGP imagined by an overconfident analyst, in which all four identifying assumptions in \cite{angrist1996} are embraced. Unbeknownst to the analyst, the monotonicity assumption is in fact violated. Helpfully, when asked to estimate bounds, our Algorithm~\ref{alg:esharp_bound} reports that the causal query is \textit{infeasible}. Recall that the true DGP corresponds to model (b), in which defiers are present; because the algorithm fails to locate any DGPs in which the observed information is consistent with the absence of defiers, it provides a clear warning to users that the assumption cannot be defended. However, if the analyst na\"ively applied the traditional instrumental variables two-stage least squares estimator, they would not be alerted to this fact. Rather, they would obtain a point estimate of $-$0.74, roughly twice the true LATE. Put differently, \textit{the standard IV approach ignores observable implications of underlying assumptions}. In contrast, our algorithm flags faulty theory by identifying infeasible scenarios, forestalling fruitless inquiry.

\subsection{Coverage of Confidence Bounds}\label{sub:coverage}

\begin{figure}[!ht]
  \centering
   \caption{\textbf{Coverage of confidence bounds.} Each of 1,000 simulations is depicted with a horizontal line. For each simulation, a horizontal error bar represents a 95\% confidence bound obtained per Section~\ref{sec:inference}. All confidence bounds fully contain the population bounds,  indicating 100\% coverage. The upper (lower) row of panels reflect confidence bounds obtained with the Bernoulli-KL (asymptotic) method. Columns of panels report confidence bounds obtained using samples of various sizes. Vertical dotted white lines show true population lower and upper bounds, which contain the true ATE of $-0.25$; vertical dashed black lines indicate zero. 
   }
   \includegraphics[width = .7\textwidth]{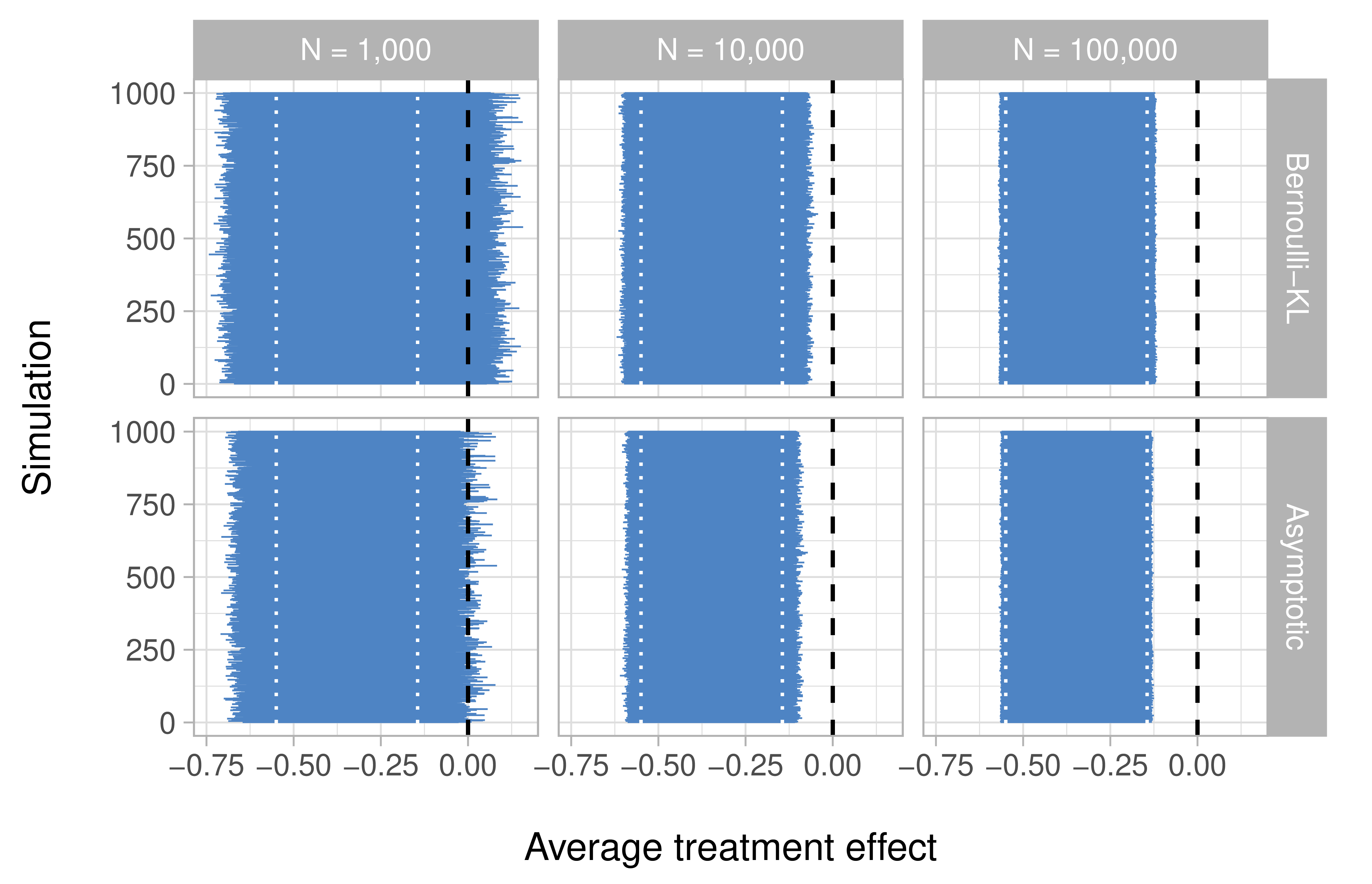}
   \label{fig:coverage}
\end{figure}

In applied settings, the bounds estimated by our algorithm will be subject to sampling error. We now evaluate the performance of confidence bounds that characterize this uncertainty, constructed according to Section~\ref{sec:inference}, using the instrumental variable model of Figure~\ref{fig:iv}(b). Specifically, we draw samples of $N=1,000$, $N=10,000$, or $N=100,000$ observations from this DGP. For each sample, we then compute estimates of eight quantities: $\Pr(Z_i=z, X_i=x, Y_i=y)$ for all $x, y, z \in \{0, 1\}$. These quantities form the basis of estimated bounds, by the plug-in principle. To quantify uncertainty, we compute 95\% confidence regions on the same observed quantities, then convert them to polynomial constraints for inclusion in Algorithm~\ref{alg:esharp_bound}. Optimizing subject to these confidence constraints produces confidence bounds, depicted in Figure~\ref{fig:coverage}. For each combination of sample size and uncertainty method, we draw 1,000 simulated datasets and run Algorithm~\ref{alg:esharp_bound} once.

Table~\ref{tab:bias} reports average values of estimated lower (upper) confidence bounds obtained by Algorithm~\ref{alg:esharp_bound} over 1,000 simulated datasets, for varying $N$. At all sample sizes, estimated bounds are centered on population bounds. Figure~\ref{fig:uncertainty} shows confidence bounds obtained across methods and sample sizes. The Bernoulli-KL method produces wider confidence intervals at all $N$; at $N=1,000$, it is generally unable to reject zero, whereas the asymptotic method does so occasionally. Differences in interval width persist but shrink rapidly as sample size grows and both methods collapse on population bounds. As discussed in Section~\ref{sec:inference}, we find more conservative coverage for confidence bounds on the ATE (100\% coverage of population bounds), compared to coverage of the underlying confidence regions on the observed quantities (95\% joint coverage of observed population quantities for the asymptotic method).

% latex table generated in R 3.6.3 by xtable 1.8-4 package
% Tue Sep 14 21:23:50 2021
\begin{table}[!ht]
\caption{\textbf{Bias of estimated bounds.} Average lower (upper) estimated bounds simulated datasets of varying size. Average estimated bounds correspond closely to population bounds.}
\label{tab:bias}
\centering
\begin{tabular}{l|rrr|r}
  \hline
Quantity & $N=1,000$ & $N=10,000$ & $N=100,000$ & Population \\ 
  \hline
Lower bound & $-$0.549 & $-$0.551 & $-$0.551 & $-$0.550 \\ 
Upper bound & $-$0.144 & $-$0.146 & $-$0.146 & $-$0.146 \\ 
   \hline
\end{tabular}
\end{table}

\subsection{More Complex Bounding Problems}\label{sub:gen_sim}

We now examine four hypothetical DGPs, shown in Figure \ref{fig:common}, featuring various threats to inference. Throughout, we target the ATE of $X$ on $Y$. Panel (a) illustrates outcome-based selection: we
observe unit $i$ only if $S_i=1$, where $S_i$ may be affected by
$Y_i$. Selection severity, $\Pr(S_i=0)$, is known, but no information about $\Pr(X_i=x, Y_i=y | S_i=0)$ is available. $X_i$ and $Y_i$ are also confounded by unobserved $U_i$. Bounding in this setting is a nonlinear program, with an analytic solution recently derived in \citet{garbiel2020}. Panel (b) illustrates measurement error: an unobserved confounder $U_i$ jointly causes $Y_i$ and its proxy $Y_i^\ast$, but only treatment and the proxy outcome are observed. Bounding in this setting is a linear problem. A number of results for linear measurement error were recently presented in \citet{finkelstein2020partial}; here, we examine the monotonic errors case, where $Y_i^\ast(Y_i=1) \ge Y_i^\ast(Y_i=0)$. Panel (c) depicts missingness in outcomes, i.e.\ nonresponse or attrition. Here, $X_i$ affects both the partially observed $Y_i$ and response indicator $R_i$; if $R_i=1$, then $Y_i^\ast=Y_i$, but if $R_i=0$, then $Y^\ast_i$ takes on the missing value indicator \texttt{NA}. Nonresponse on $Y_i$ is differentially affected by both $X_i$ and the value of $Y_i$ itself (i.e.\ ``missingness not at random,'' MNAR); \citet{manski1990} provides analytic bounds. Finally, panel (d) depicts joint missingness in both treatment and outcome---sometimes a challenge in longitudinal studies with dropout---with MNAR on $Y_i$.

Figure~\ref{fig:common_results}(a--c) illustrates how Algorithm~\ref{alg:esharp_bound} recovers sharp bounds. Each panel shows progress in time, converging on known analytic results depicted at the right of each plot. Primal bounds (blue) widen over time as more extreme, observationally equivalent models are found. Dual bounds (red) narrow as the outer envelope is tightened. When a region cannot possibly produce a more extreme value than a previously discovered primal point, it is eliminated from consideration. Optimization proceeds by simultaneously searching for more extreme primal points and narrowing the dual envelope. Analysts can terminate the process at any time, reporting guaranteed-valid dual bounds along with their worst-case suboptimality factor, $\varepsilon$---or await complete sharpness, $\varepsilon=0$. 

\begin{figure}[p]
  \centering
   \caption{\textbf{Various threats to
     inference.} Panels depict (a) outcome-based selection, (b) measurement
   error, (c) nonresponse and (d) joint missingness. In each graph, $X$ and $Y$ are treatment and
   outcome, respectively. Dotted red regions represent observed information. In (a),
   the box around $S$ indicates selection: other variables are only
   observed conditional on $S=1$. In (b), $Y^\ast$ represents a
   mismeasured version of the unobserved true $Y$. In (c), $R_Y$ indicates
   reporting, so that $Y^\ast=Y$ if $R=1$ and is missing otherwise. In (d), both treatment and outcome can be missing; and missingness on $X$ can affect missingness on $Y$.}
   \label{fig:common}
  \vspace{-1ex}
  \begin{minipage}[b]{.45\textwidth}
 \centering
 \begin{tikzpicture}%[node distance=.5 cm and .5 cm]
 \node (U) at (1.5,1) {$U$};
 \node (X) at (0.5,0) {$X$};
 \node (Y) at (2.5,0) {$Y$};
 \node[draw] (S) at (4.5,0) {$S$};
 \path[dashed] (U) edge [->](X);
 \path[dashed] (U) edge [->](Y);
 \path (X) edge [->](Y);
 \path (Y) edge [->](S);
 \draw [rotate around={0:(0,0)}, -, dotted, red, semithick, rounded corners] (.1,-.4) rectangle (2.9,.4);
\node (a) at (-.95, -.5) {(a)};
\draw (-1.25,-.75) rectangle (-.65, -.2);
\draw (-1.25,-.75) rectangle (5.75,1.75);
%\node (a) at (-.95, -1) {(a)};
%\draw (-1.25,-1.25) rectangle (-.65, -.7);
%\draw (-1.25,-1.25) rectangle (5.75,1.75);
%\useasboundingbox (-1.25,-1.25) rectangle (5.75,1.75);
%\useasboundingbox (-1.25,-1.25) rectangle (5.5,1.5);
 \end{tikzpicture}
 \end{minipage}
 \begin{minipage}[b]{.45\textwidth}
 \centering
 \begin{tikzpicture}%[node distance=.5 cm and .5 cm]
 \node (U) at (3,0) {$U$};
 \node (X) at (1,0) {$X$};
 \node (Ya) at (2,1) {$Y$};
 \node (Yb) at (4,1) {$Y^\ast$};
 \path[dashed] (U) edge [->](Ya);
 \path[dashed] (U) edge [->](Yb);
 \path (X) edge [->](Ya);
 \path (Ya) edge [->](Yb);
 \draw [rotate around={0:(0,0)}, -, dotted, red, semithick, rounded corners] (.6,-.4) rectangle (1.4,.4);
 \draw [rotate around={0:(0,0)}, -, dotted, red, semithick, rounded corners] (3.6,.6) rectangle (4.4,1.4);
\node (b) at (-.95, -.5) {(b)};
\draw (-1.25,-.75) rectangle (-.65, -.2);
\draw (-1.25,-.75) rectangle (5.75,1.75);
%\node (b) at (-.95, -1) {(b)};
%\draw (-1.25,-1.25) rectangle (-.65, -.7);
%\draw (-1.25,-1.25) rectangle (5.75,1.75);
%\useasboundingbox (-1.25,-1.25) rectangle (5.75,1.75);
 \end{tikzpicture}
 \end{minipage}\\[1ex]
 \begin{minipage}[b]{.45\textwidth}
 \centering
 \begin{tikzpicture}%[node distance=.5 cm and .5 cm]
 \node (X) at (0,0) {$X$};
 \node (Ya) at (1,1) {$Y$};
 \node (R) at (3,1) {$R_Y$};
 \node (Yb) at (5,1) {$Y^\ast$};
 \path (X) edge [->](Ya);
 \path (X) edge [->, bend right=20](R);
 \path (Ya) edge [->](R);
 \path (R) edge [->](Yb);
 \path (Ya) edge [->, bend left=20](Yb);
 \draw [rotate around={0:(0,0)}, -, dotted, red, semithick, rounded corners] (-.4,-.4) rectangle (.4,.4);
 \draw [rotate around={0:(0,0)}, -, dotted, red, semithick, rounded corners] (2.6,.6) rectangle (5.4,1.4);
\node (c) at (-.95, -.5) {(c)};
\draw (-1.25,-.75) rectangle (-.65, -.2);
\draw (-1.25,-.75) rectangle (5.75,1.75);
%\node (c) at (-.95, -1) {(c)};
%\draw (-1.25,-1.25) rectangle (-.65, -.7);
%\draw (-1.25,-1.25) rectangle (5.75,1.75);
%\useasboundingbox (-1.25,-1.25) rectangle (5.75,1.75);
 \end{tikzpicture}
 \end{minipage}
 \begin{minipage}[b]{.45\textwidth}
 \centering
 \begin{tikzpicture}%[node distance=.5 cm and .5 cm]
 \node (Xa) at (0,0) {$X$};
 \node (RX) at (2,0) {$R_X$};
 \node (Xb) at (4,0) {$X^\ast$};
 \node (Ya) at (1,1) {$Y$};
 \node (RY) at (3,1) {$R_Y$};
 \node (Yb) at (5,1) {$Y^\ast$};
 \path (Xa) edge [->](Ya);
 \path (Ya) edge [->](RY);
 \path (RX) edge [->](RY);
 \path (RX) edge [->](Xb);
 \path (Xa) edge [->, bend right=20](Xb);
 \path (RY) edge [->](Yb);
 \path (Ya) edge [->, bend left=20](Yb);
 \draw [rotate around={0:(0,0)}, -, dotted, red, semithick, rounded corners] (1.6,-.4) rectangle (5.4,1.4);
% \draw [rotate around={0:(0,0)}, -, dotted, red, semithick, rounded corners] (1.2,-.4) -- (2.6,1.4) -- (5.6,1.4) -- (4.2,-.4) -- cycle;
\node (d) at (-.95, -.5) {(d)};
\draw (-1.25,-.75) rectangle (-.65, -.2);
\draw (-1.25,-.75) rectangle (5.75,1.75);
%\node (d) at (-.95, -1) {(d)};
%\draw (-1.25,-1.25) rectangle (-.65, -.7);
%\draw (-1.25,-1.25) rectangle (5.75,1.75);
%\useasboundingbox (-1.25,-1.25) rectangle (5.75,1.75);
 \end{tikzpicture}
 \end{minipage}
\end{figure}
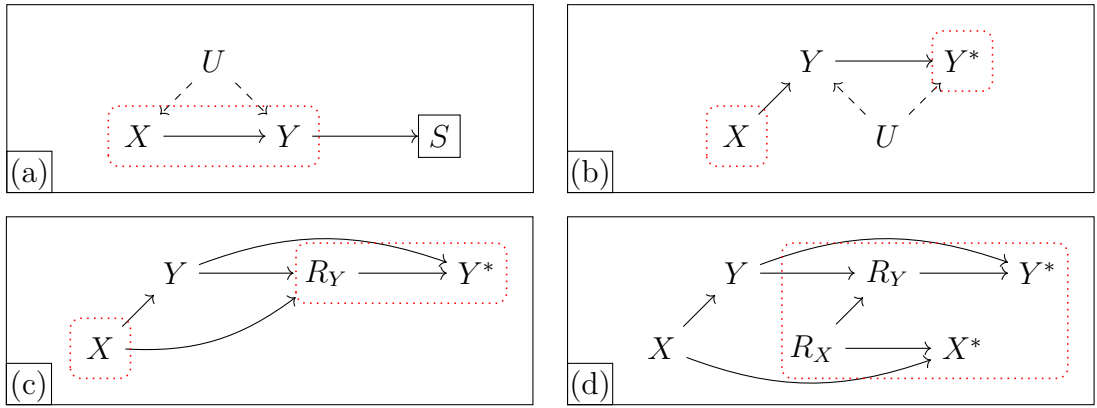
\begin{figure}[p] %
   \caption{\textbf{Computation of ATE bounds.} Progress of Algorithm~\ref{alg:esharp_bound} in simulated Figure~\ref{fig:common}(a--d) DGPs. Black error bars are known analytic bounds, $y$-axes are ATE values, and $x$-axes are runtimes of Algorithm~\ref{alg:esharp_bound}. Prior analytic bounds are sharp for settings (a--c). In setting (d), Algorithm~\ref{alg:esharp_bound} achieves point identification, but \citet{manski1990} bounds do not. Red regions are dual bounds, which always contain sharp bounds and the unknown true causal effect; these can only narrow over time, converging on optimality. Blue regions are primal bounds, which can only widen over time as more extreme models are found. Optimization stops when primal and dual bounds meet, indicating bounds are sharp.}
   \label{fig:common_results}
   \centering
        \begin{subfigure}[b]{.45\textwidth} 
        \centering 
        \subcaption{Outcome-based selection}
        \includegraphics[width=5.5cm]{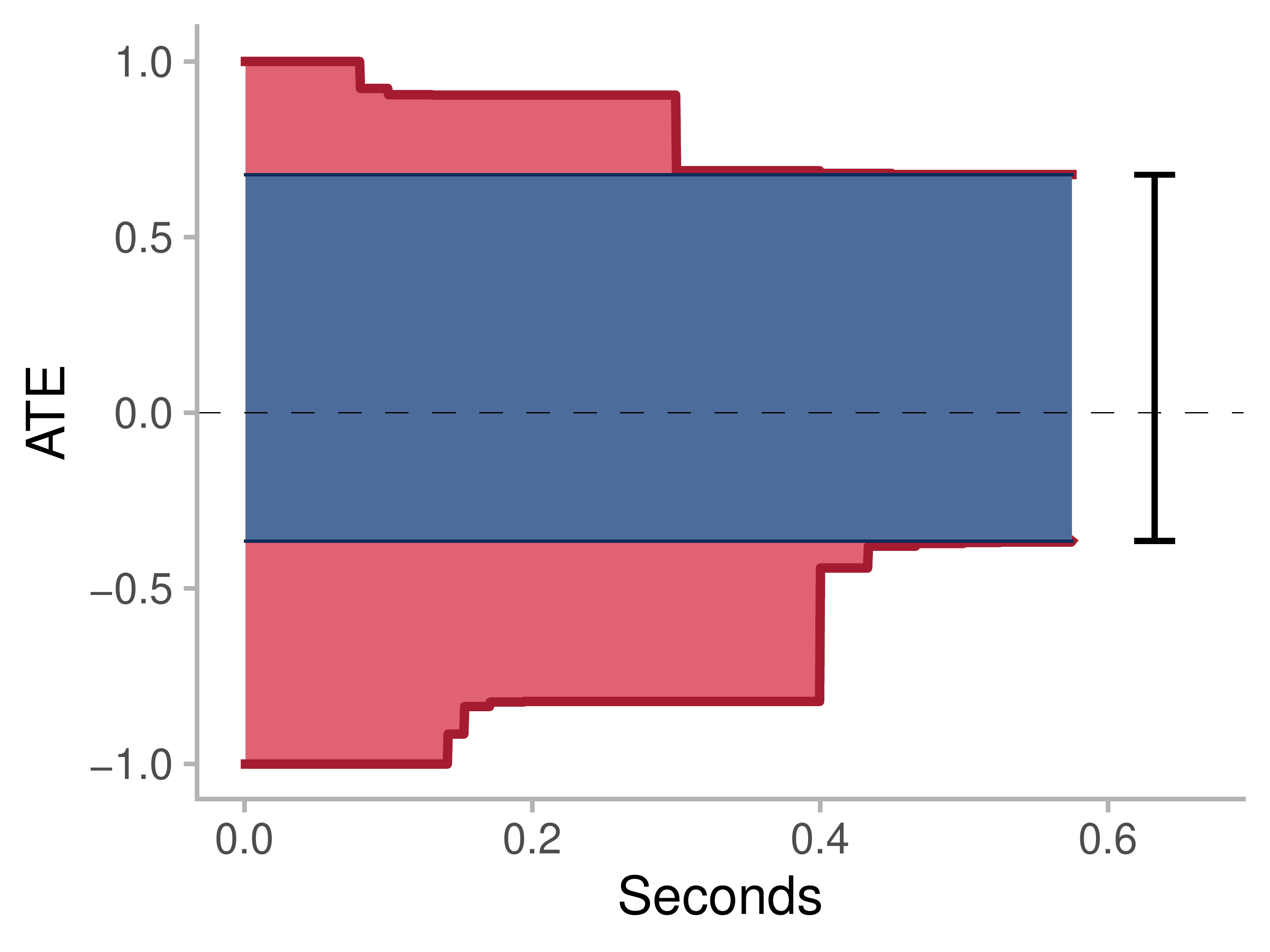}
        \label{subfig:selection} 
        \end{subfigure} %
        \begin{subfigure}[b]{.45\textwidth}
        \centering      
        \subcaption{Measurement error}
        \includegraphics[width=5.5cm]{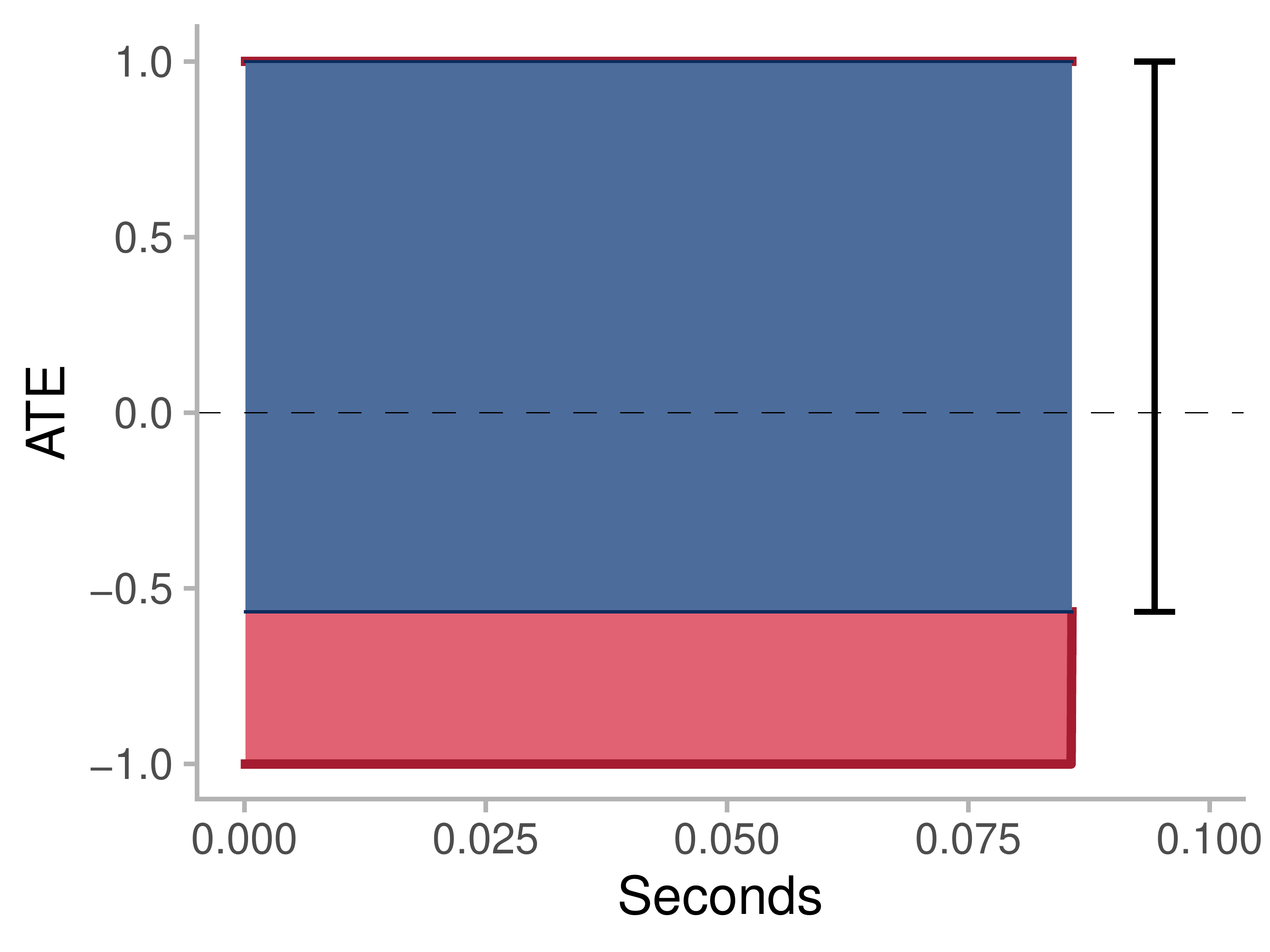}
 \label{subfig:measurement} 
        \end{subfigure} %
        \begin{subfigure}[b]{.45\textwidth}
        \centering 
      \subcaption{Nonresponse}
        \includegraphics[width=5.5cm]{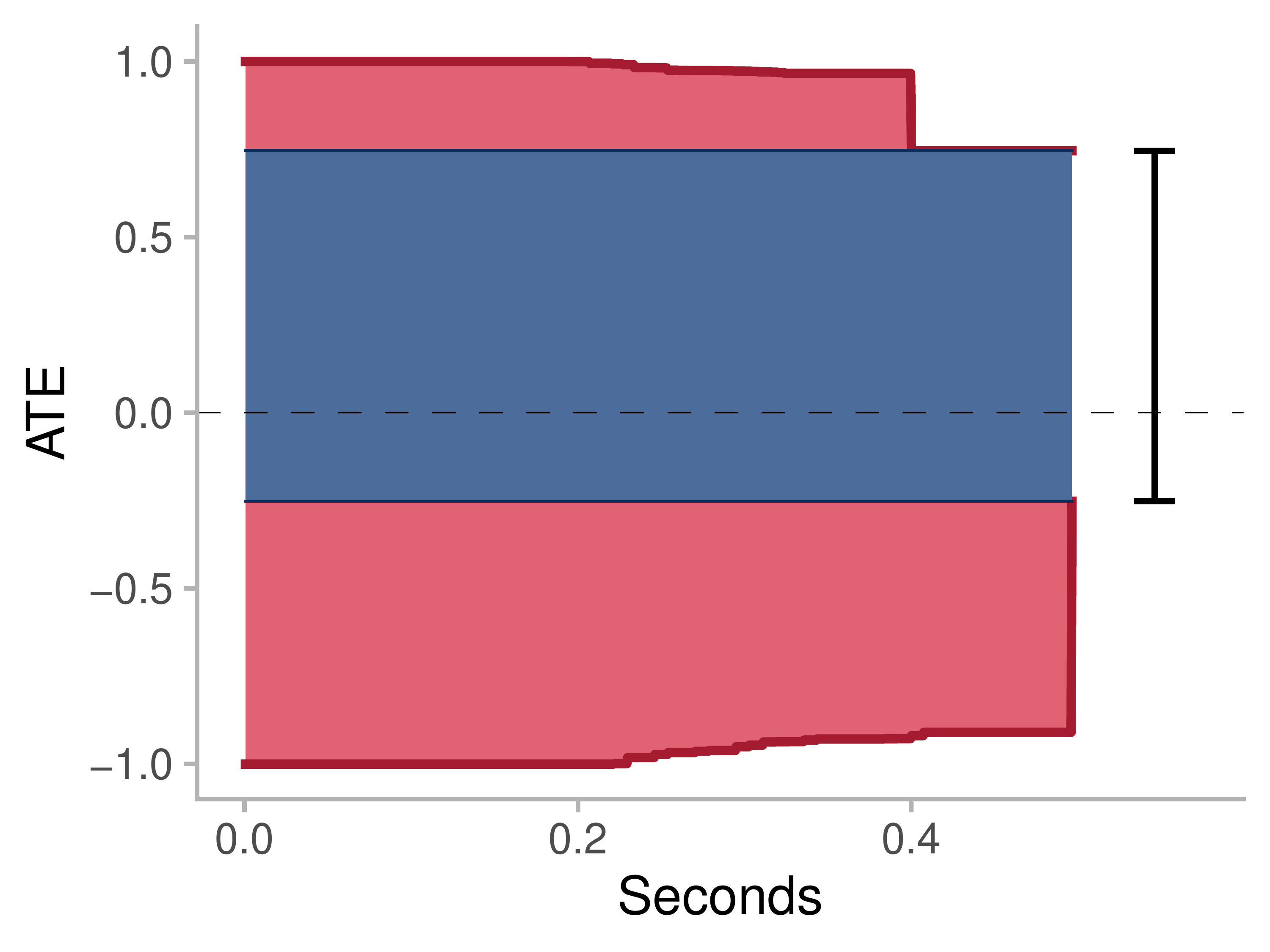}
     \label{subfig:nonresponse} 
        \end{subfigure} %
       \begin{subfigure}[b]{.45\textwidth}
        \centering 
      \subcaption{Joint missingness}
        \includegraphics[width=5.5cm]{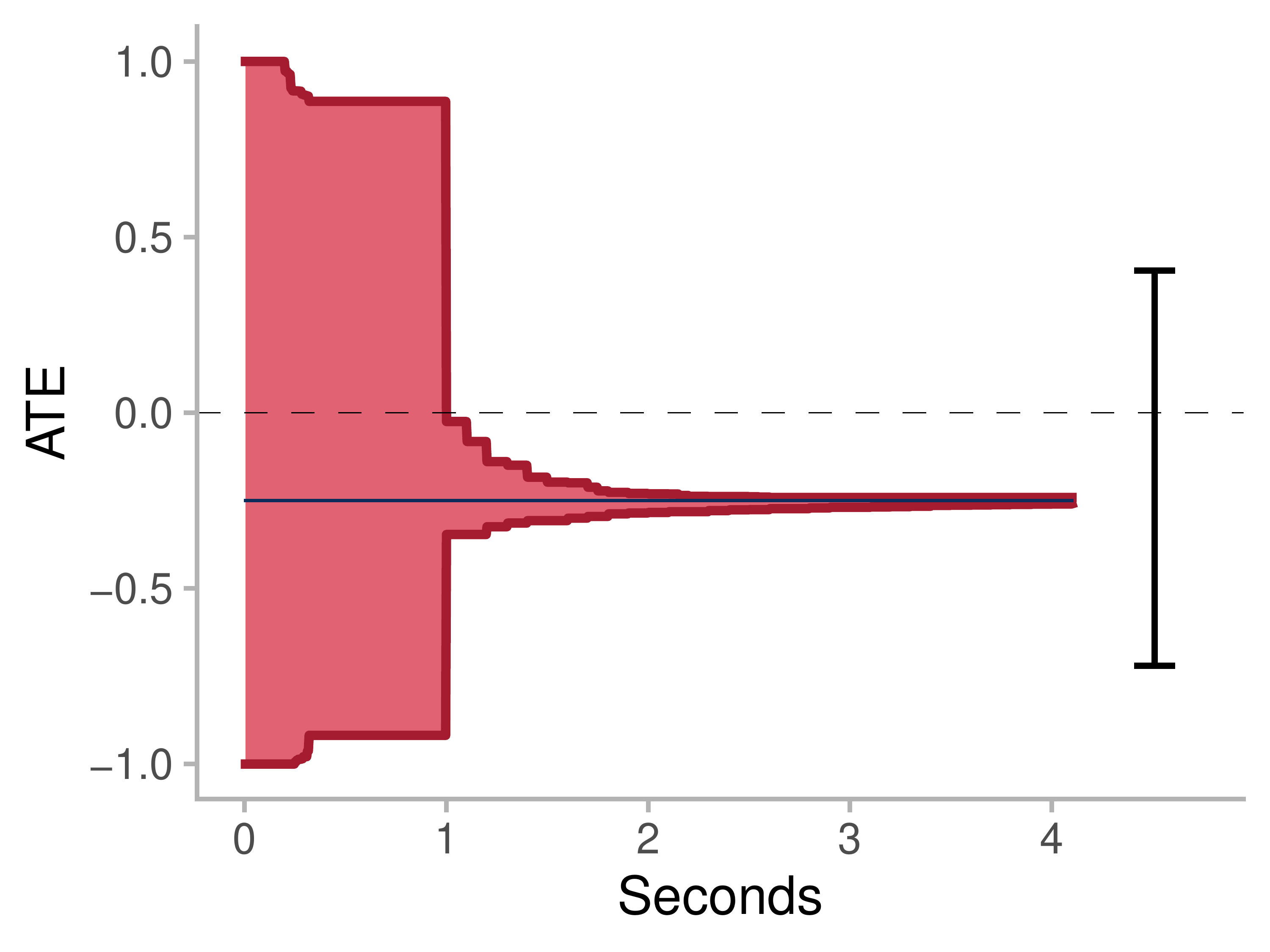}
     \label{subfig:joint_missingness} 
        \end{subfigure} %
        \vspace{-2ex}
    \end{figure}
\clearpage

In Figure~\ref{fig:common_results}(a--c), the algorithm converges on known analytic results. Ultimately, in the selection simulation (a), Algorithm~\ref{alg:esharp_bound} achieves bounds of $[-0.37, 0.68]$, correctly recovering \citeauthor{garbiel2020}'s (\citeyear{garbiel2020}) bounds; in (b), measurement error bounds are $[-0.57, 1.00]$, matching \citet{finkelstein2020partial}; and in (c), outcome missingness bounds are $[-0.25, 0.75]$, equaling \citet{manski1990} bounds. Somewhat counterintuitively, Figure~\ref{fig:common_results}(d) shows dual bounds collapsing to a point, correctly point-identifying the ATE at $-0.25$ despite severe missingness. This surprising result turns out to be a special case of an approach using ``shadow variables'' recently developed by \citet{miao2015identification}.\footnote{Specifically, it can be shown the ATE is identified for the Figure~\ref{fig:common}(d) graph only among faithful distributions where $X \to Y$ is non-null---i.e.\ almost everywhere in the model space.} This example illustrates that the algorithm is general enough to recover results even when they are not widely known in a particular model; note that the commonly used approach of \citet{manski1990} produces far looser bounds of $[-0.72, 0.40]$, failing to exploit causal structure given in Figure~\ref{fig:common}(d). This result suggests our approach enables an empirical investigation of complex models where general identification results are not yet available. Situations where bounds converge suggest models where point identification via an explicit functional may be possible, potentially enabling new identification theory.

\section{Potential Critiques of the Approach}\label{sec:critique}
Below, we briefly discuss several potential critiques of our method.\\[-2.5ex]

\noindent\textbf{``The user must know the true causal model.''}\\
Our algorithm requires users specify a causal graph and assumptions, but in many applications, the true DGP is unknown. This is precisely the obstacle that motivates our approach, which allows for valid inferences in the \textit{absence} of complete information. Rather than assert a faulty ``complete'' model, the user need only input what they know or believe. The algorithm then outputs the most precise possible solution given that information; key assumptions can be relaxed further using easily incorporated sensitivity analyses, as needed. We note the difficulty of declaring a causal theory, even a partial one, is universal: any attempt to draw causal inferences from data---even in experimental settings---is premised (often implicitly) on underlying causal theory. Making assumptions explicit is not a trade-off relative to other methods, but a boon for research transparency.\\[-2.5ex]

\noindent\textbf{``The bounds may be too wide to be informative.''}\\
Yes.

When a point-identified solution exists, our algorithm will discover it. As Section~\ref{sub:gen_sim} shows, this can occur in surprising scenarios and may help reveal new identification theory. However, when point-identification is impossible, our approach produces sharp bounds. These bounds may be insufficient for an analyst to achieve a goal such as discerning the sign of a causal effect. This is simply a fact about the limitations of the research design---as we prove, it is impossible to narrow the bounds further without additional information. Again, there is no tradeoff: incorrect point estimates based on faulty assumptions are also uninformative. When sharp bounds incorporating all defensible assumptions are wide, it means progress will require collecting more data or justifying additional assumptions. \\ [-2.5ex]
%

%In addition, because our algorithm is fully modular, analysts have the ability to add and remove identifying assumptions in order to gauge what additional conditions would need to be present to narrow the range of possible solutions even further. Such exploratory analyses can form the bases of useful debates and targeted follow-up studies.\\

\noindent\textbf{``What about continuous variables?''}\\
Our approach applies to discrete data, but analysis of continuous variables can often still proceed with some adjustments. Discrete approximations often suffice in applied work. \citep[Indeed, ``all data as observed are discrete,'' ][p.\ 133]{Rubin1981}. When continuous treatments (e.g.\ birth date, vehicle speed) often affect discrete outcomes (school admittance, police stops) only when exceeding a threshold, discretization is lossless. Moreover, when analyzing discrete treatments and continuous outcomes, much of our theory generalizes to estimands involving expectations of the outcome. Future work may study our method's applicability to bounded continuous variables with smooth effects.\\ [-2.5ex]

\noindent\textbf{``The bounds will take too long to compute.''}\\
Computation time for sharp bounds may sometimes be prohibitive, but our approach is likely still faster than manual derivation. Notably, the algorithm recovers several recently published analytic results in mere seconds \citep{garbiel2020,miao2015identification,KnoxLoweMummolo2020}. Second, when computation time is long, our algorithm's ``anytime'' guarantee ensures premature termination will still produce valid bounds and report a worst-case looseness factor for the resulting non-sharp bounds. 

\section{Future Work with Automated Bounding}\label{sec:discussion}

Causal inference is a central goal of science, and several established techniques can estimate causal quantities under ideal conditions. But in many applications, these conditions are simply not satisfied, and developing new analytic solutions is often intractable. For knowledge accumulation to proceed in the messy world of applied statistics, a general solution is needed. We present a tool to automatically produce sharp bounds on causal quantities in settings involving discrete data. Our approach involves a reduction of all such causal queries to polynomial programming problems, enables efficient search over observationally indistinguishable DGPs, and produces sharp bounds on arbitrary causal estimands. This approach is sufficiently general to accommodate a range of classic inferential obstacles.

Beyond providing a general tool for causal inference, our approach aligns closely with recent calls to improve research transparency by requiring the explicit declaration of estimands, identifying assumptions, and theory \citep{Miguel2014,lundbergetal2021}. With a common understanding of goals and premises, scholars can have meaningful debates over the credibility of research. When aspects of a theory are contested, our approach allows for a fully modular exploration of how assumptions shape empirical conclusions. Scholars can learn whether a particular assumption is empirically consequential, and if so, craft a targeted line of inquiry to probe its validity. Our approach can also act as a safeguard for analysts, flagging assumptions as infeasible when they conflict with observed information. This means hopeless research projects can be abandoned before wasting effort or disseminating untruths. 

Future work should seek to reduce computation time for sharp bounds, especially when incorporating point-identified subquantities or additional semi-parametric modeling approaches. Causal inference scholars may also use this method as an exploratory tool to aid in the discovery of new identification theory. These lines of inquiry now represent the major open questions in discrete causal inference.

%When faced with unavoidable obstacles to point identification, researchers now have a general and automated method to recover the most precise possible answer to a causal query given the information at hand.

 %These speed issues are what remain of major open questions in discrete causal inference.

%The central limitation of our approach is that it is limited to problems involving discrete data. However, we note that discrete data describe an immense range of settings, including cases where the research goal allows for the coarsening of continuous variables (e.g., the analyst is only interested in learning how events above some threshold affect an outcome). Still, future work can now turn to the task of developing similar techniques for continuous settings, perhaps by XXX [suggestion for how this might be achieved?] XXX. \\

%Note: discuss categorical outcomes somewhere in main text\\

%- applied causal inference is very much a case by case endeavor.

%- we now know when quantities of interest are point identified. but conditions often deviate.

%- if not point identified, then what? in some very simple cases we had bounds, and sometimes we even know they're sharp

%- but even these are only applicable under ideal/very simple deviations. very simple causal structures, one problem at a time (wall of cites for special cases).

%- real world is often messier in many ways (causal structure, multiple problems). but scientists have a lot of knowledge about how things like missingness, selection, etc happen (can incorporate this through assumptions)

%- brief recap of our solution: xxx
\bibliographystyle{chicago}

\singlespacing

\bibliography{biblio}

\clearpage
\doublespacing
\appendix

\section{Examples, Algorithms, and Detailed Discussion}
\label{app:algorithms}
\setcounter{equation}{0}

\subsection{Canonicalization of DAGs}
\label{app:canonical_dag}

In this appendix, we summarize the process for obtaining a canonical hidden variable DAG, presented as Definition 4.6 in \cite{evans2016graphs}. Theorem 4.13 in \cite{evans2016graphs} shows that the marginal model of any hidden variable DAG is the same as that of its canonical hidden variable DAG, and Proposition 7.4 of the same work shows that the same holds for the model for post-intervention distributions, when interventions are restricted to the main variables.

Given a hidden variable DAG $\cG$, the canonical form of the DAG is constructed by the following procedure:

\begin{enumerate}
    \item Add an edge $X_j \rightarrow X_{j'}$ for any pair of variables $X_j, X_{j'}$ such that there is a path from $X_j$ to $X_{j'}$ along which all variables between $X_j$ and $X_{j'}$ are hidden. $X_j$ and $X_{j'}$ can each be hidden or observed.
    \item Remove incoming edges to hidden variables.
    \item Remove hidden variables whose children are a subset of the children of another hidden variable.
\end{enumerate}

\noindent By construction, all latent variables in the canonical DAG will be exogenous.

\subsection{Functional Models in the Context of Determinism}
\label{app:reduced_functional_model}

The general approach for obtaining functional models for discrete hidden
variable DAGs \citep{evans2018margins, wolfe2021cardinalities} does not take
account of the kind of determinism introduced into the model by missingness
indicators, and as such may yield a functional model that is
\emph{over-parameterized}. Due to the complexity of polynomial programming, it
is beneficial to avoid excess parameters where possible. We now briefly explore
this issue.

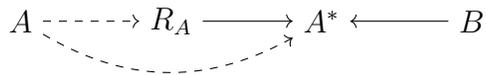
\begin{figure}[htb]
\caption{\textbf{A graph with determinism.}}
\centering
\vspace{2ex}
\label{fig:determinism-demo}
 \begin{tikzpicture}
 \node (A) at (0,0) {$A$};
 \node (B) at (2,0) {$R_A$};
 \node (C) at (4,0) {$A^\ast$};
 \node (D) at (6, 0) {$B$};
 \path[dashed] (A) edge [->] (B);
 \path[dashed, bend right=30] (A) edge [->] (C);
 \path (B) edge [->] (C);
 \path (D) edge [->] (C);
 \end{tikzpicture}
\end{figure}

Consider the scenario depicted in Figure \ref{fig:determinism-demo}. In this
graph, $A^\ast$ is a proxy for the unobserved variable $A$, which is observed with
missingness as indicated by $R_A$. When $R_A = 0$, then $A^\ast$ is
deterministically equal to a special value indicating missingness (usually denoted with the special value such as ``?'' or ``\texttt{NA}''). In addition,
$A^\ast$ is affected by $B$. This scenario might arise if $A$ is measured with
missingness \emph{and} measurement error, and the nature of the error is
affected by $B$. Of note, $A^\ast$ is not a fully deterministic function of $A$ and
$R_A$, and cannot simply be removed from the functional parameterization, as in traditional missingness without measurement error.
However, we can use the fact that it is a \emph{partially} deterministic
function of $R_A$ to reduce the number of parameters needed in the functional
model for this graph.

In general, the functional model for this graph would allocate one value of
$\epsilon_{A^\ast}$---the exogenous noise that determines $A^\ast$ in terms of its
parents---for every combination of possible responses of $A^\ast$ to its parents.
Suppose $A^\ast$ takes values in $\{0, 1, ?\}$, and $A$, $R_A$ and $B$ take values
in $\{0, 1\}$. This would correspond to $3^{8} = 6561$ possible values of
$\epsilon_{A^\ast}$. However, any such value that maps $R_A = 0$ to $A^\ast \in \{0, 1\}$ or $R_A = 1$ to $A^\ast = ?$ is
ruled out by the deterministic relationship. As a result, $\epsilon_{A^\ast}$ need
    only specify the response of $A^\ast$ in $\{0, 1\}$ to $A$ and $B$ when $R_A = 1$. This yields only $2^4 = 16$ possible values for
$\epsilon_{A^\ast}$. This example demonstrates that incorporating known deterministic relationships can
yield a non-restrictive functional parameterization with fewer parameters.

% \begin{proposition} \label{prop:functional-model} 
%   A functional model of a latent variable DAG \citep{evans2018margins, wolfe2021cardinalities} is non-restrictive of the full data law.
% \end{proposition}
% \begin{proof}
%   Extend Evans proof slightly.
% \end{proof}

\subsection{DAG Parameterization for Non-geared Graphs}

Most graphs we encounter in practice are \textit{geared}
\citep{evans2018margins}, which means they have no \textit{non-trivial
  bi-directed cycles} \cite{wolfe2021cardinalities}. When graphs are not geared,
and the target estimand as well as all empirical evidence involves only single
world probabilities, it is possible to improve the complexity of the
system. Under these circumstances, it is preferable to obtain non-restrictive
bounds on the cardinalities of latent variables according to
\cite{wolfe2021cardinalities}. All single world probabilities can be expressed
in terms of the usual DAG parameters according to the g-formula, and therefore
all functionals of such probabilities described in Corollary
\ref{cor:functional-polynomialization} can be polynomialized as well. If the
target or any of the empirical evidence involve cross-world probabilities, we
must revert to the functional model approach.

\subsection{Example of Program Simplification}
\label{app:nested_markov_example}

\begin{figure}[htb]
\caption{\textbf{A graph with conditional independence and Verma constraints.}}
\centering
\vspace{2ex}
\label{fig:nested-demo}
 \begin{tikzpicture}
 \node (A) at (0,0) {$A$};
 \node (B) at (2,0) {$B$};
 \node (C) at (4,0) {$C$};
 \node (D) at (6, 0) {$D$};
 \node (E) at (2,-2) {$E$};
 \node (F) at (4,-2) {$F$};
 \node (U1) at (4, 1) {$U_1$};
 \node (U2) at (2, -1) {$U_2$};
 \node (U3) at (6, -2) {$U_3$};
 \path[dashed] (U1) edge [->](B);
 \path[dashed] (U1) edge [->](D);
 \path[dashed] (U2) edge [->](A);
 \path[dashed] (U2) edge [->](C);
 \path[dashed] (U2) edge [->](E);
 \path[dashed] (U3) edge [->](D);
 \path[dashed] (U3) edge [->](F);
 \path (A) edge [->](B);
 \path (B) edge [->](C);
 \path (C) edge [->](D);
 \path (A) edge [->](E);
 \path (E) edge [->](C);
 \path (D) edge [->](F);
 \end{tikzpicture}
\end{figure}
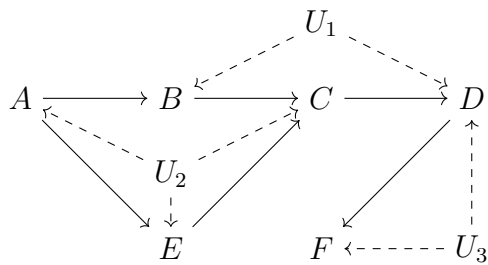

Consider the graph presented in Figure \ref{fig:nested-demo}. We will use this graph to illustrate a number of points raised in the main body of the paper. Suppose we are interested in the ATE of $E$ on $C$. First, we will explicitly construct the functional model of this graph, then use it to generate a simple polynomial program that bounds a causal target. Next, we will employ several of the strategies described in Section \ref{sec:simp} to simplify the program, demonstrating the importance of these strategies in obtaining tractable program formulations. Finally, we will observe that a broader class of partial identification problems than previously recognized can be formulated as linear programs.

Suppose all observed variables in the graph above are binary. In constructing a functional model, we first note that $U_2$ is responsible for determining the values of $A$, $C$ and $E$ in response to their parents. $A$ has no parents, $E$ has one parent, and $C$ has two parents. Therefore $U_2$ takes values in a state space of size $2^1 \times 2^2 \times 2^4 = 128$. Next, we suppose $U_1$ is responsible for determining the value of $B$ in response to $A$, and therefore has size $2^2 = 4$. $U_3$ is left to determine the value of $F$ in response to $D$, and of $D$ in response to $U_1$ and $C$. It therefore takes values in space of size $2^8 \times 2^2 = 1024$.\footnote{It is also possible to construct a functional model by first taking $U_3$ to be responsible for determining $F$ in response to $D$, and then $U_1$ to be responsible for determining $B$ in response to $A$ and $D$ in response to $C$ and $U_3$. By a simple symmetry argument, the two functional models yield the same number of parameters.}

To construct the polynomial program, we begin with the non-negativity and linear marginalization constraints on the parameters of the distributions of the disturbances (for simplicity, we abstain from eliminating one parameter per distribution using the sum-to-unity constraint):
\begin{align*}
    &\Pr(U_i = u) \ge 0 &\forall i, \forall u \in \Omega_{U_i}\\
    \sum_{u \in \Omega_{U_i}} &\Pr(U_i = u)=1 & \forall i.
\end{align*}

We then add constraints encoding the empirical evidence $\cE$. For simplicity,
we assume that we observe the full joint distribution $\Pr(A = a, B = b, C = c, D = d, E = e, F = f)$,
which is a vector of size $2^6 = 64$, corresponding to $64$ equality constraints
in the program. There are $3$ disturbance variables in this graph, including
$\epsilon_E$, leading to polynomials in these equality constraints with terms of
degree $3$. Given the cardinalities of the disturbances, there are $2^4 \times
2^7 \times 2^{10}= 2,097,152$ possible combinations of disturbance assignments. By a
simple exchangeability argument, the same number of possible combinations lead
to each outcome in the state space. As there are $2^6$ outcomes,
each of the 64 polynomial equality constraints for $\cE$ will have
$\frac{2^{21}}{2^6} = 2^{15}$ terms, again each of degree $3$. This is a very
large program.
\begin{align}
  \Pr(A = a, B = b, C = c, D = d, E = e, F = f) = \sum_{u \in \Omega_U} \prod_i \Pr(U_i = u)
  \bbone(u \implies a, b, c, d, e, f)
\end{align} 

We now consider the strategies described in Section~\ref{sec:simp}. First, observe that there are only 31 nested Markov parameters for this graph, corresponding to 31 polynomial equality constraints encoding $\cE$: a substantial savings over the $64$ parameters of the na\"ive parameterization. This reduced parameterization is possible because it encodes standard conditional independences, such as $F \perp A \mid D$. In addition, it encodes Verma constraints, which emerge either (i) from independences in post-intervention distributions or (ii) from the irrelevance of an intervention to a particular distribution. In this case, $A \perp \{D, F\} \mid do(C)$. As discussed in the main text, each equality constraint can be used to reduce the number of parameters needed in a non-restrictive reduction that can express every possible distribution in the model.

Recall that each nested Markov parameter corresponds to the identified probability of a single world event, where the event is specified in terms of variables in a single district, and the intervention is on all parents of the district relevant to those variables. For example, in this case, one of the nested Markov parameters is $\Pr\big(b=1,f=1|d=1, do(a=1,c=1)\big)$. We can now make use of Proposition~\ref{prop:relevant-latents} to reason that each of these polynomial constraints must involve only disturbances from a single district. Therefore in the equations corresponding to nested Markov parameters for the district corresponding to $U_2$, parameters of the distributions of $U_1$ and $U_3$ will all sum out, and we will be left with equations that are linear in the parameters of $U_2$. Likewise, in equations corresponding to nested Markov parameters for the district containing descendants of $U_1$ and $U_3$, parameters for the distribution of $U_2$ will factor out, and we will be left with a quadratic equation.

Finally, we can make use of Proposition \ref{prop:constraint-complexity} to note that constraints involving nested Markov parameters corresponding to the $\{U_1, U_3\}$ district can be dropped from the program. This is because they only involve parameters for the distributions of $U_1$ and $U_3$, which do not appear in any constraint involving parameters for the distribution of $U_2$. The target, by contrast, involves only parameters for the distribution of $U_2$.

As a result of taking the three steps described in Section \ref{sec:simp}, we have taken this problem from a \emph{polynomial} program involving $1156$ parameters to a \emph{linear} program involving only $2^7 = 128$ parameters and fewer constraints. This example also motivates the following corollary, which expands the class of partial identification problems that can be formulated as linear programs relative to known results \citep{balke1997bounds, finkelstein2020partial, wolfe2019inflation}.

\begin{corollary}
  Suppose $\cG$ is a hidden variable DAG with observed variables $\bV$, $\cC = \{ V_{\ell}(\ba_{\ell}) =
  v_{\ell} \mid \ell \in \mathcal L\}$ is a set of counterfactual statements, and $\Pr(\cC)$ is the target of interest. Further suppose that the full joint distribution $\Pr(\bV=\bv)$ is observed. Then $\Pr(\cC)$ can be sharply bounded given the observed data by optimizing a linear program if all $\{V_\ell | \ell \in \mathcal L\}$ are in the same single-latent-variable district.
\end{corollary}
\begin{proof}
  Because the common district of $\cC$ contains only a single latent variable, by Proposition \ref{prop:relevant-latents} the objective will be linear in the parameters of the distribution of that latent variable. By Proposition \ref{prop:constraint-complexity}, the constraints will not involved parameters corresponding to other districts. By Algorithm \ref{alg:polynomial-program-simple}, no single term in a constraint will involve multiple parameters for the same latent distribution, meaning that all constraints involving only parameters corresponding to a single-variable district will be linear. The non-negativity and sum-to-unity constraints on the parameters of the latent-variable distribution are also linear. It follows that the objective and all constraints are linear.
\end{proof}

\subsection{Constructing the Polynomial Program}

Algorithm~\ref{alg:polynomial-program-simple} constructs a polynomial program
to sharply bound any factual or counterfactual target of inference, $\cT$, that
is a polynomial fraction or monotonic transformation thereof. In addition to $\cT$, the algorithm takes as input a possibly non-canonical DAG $\cG$; empirical evidence $\cE$, modeling assumptions $\cA$, and sample space of possible outcomes for the main variables, $\cS(\bV)$. It
produces an optimization problem with a polynomial objective subject to
polynomial constraints. This polynomial programming problem is equivalent to the
original causal bounding problem. 

\begin{algorithm}[htb!]
  %\thisfloatpagestyle{empty}
\caption{~Constructing a Polynomial Program}
\label{alg:polynomial-program-simple}
\small
\vspace{1ex}
%% \hspace*{\algorithmicindent}
\begin{tabular}{ll}
  \textbf{Input:} & graph $\cG$, evidence $\cE$, assumptions $\cA$, sample space
  $\cS(\bV)$, target $\cT$ \\
  \textbf{Output:} & polynomial program in parameters $\cP_{\bU}$ or $\cP_{\bU} \cup s$
\end{tabular}
%% \hspace*{\algorithmicindent} 
\begin{algorithmic}[1]
  \vspace{1ex}
  \Statex \textit{Initialization}
  \State initialize empty constraint set $\cC \leftarrow \varnothing$
  \State $\cG \leftarrow$ canonicalize $\cG$
  \State $\cP_{\bU} \leftarrow$ parameters of functional model for $\cG$
  \vspace{1ex}  
  \Statex \textit{Polynomialize objective function}  
  \State $\cT \leftarrow$ polynomial-fractionalize($\cT$) 
  \If{$\cT$ contains fractions}
  \State polynomialize($\cT = s$) and append to $\cC$
  \State $\cT \leftarrow s$
  \EndIf
  \vspace{1ex}
  \Statex \textit{Polynomialize constraints}    
  \For{$\big( g(\cP_{\bV})~\bstar~\alpha \big) \in \big( \mathcal E \cup \mathcal A \big)$}
  \State polynomialize$\big( g(\cP_{\bV})~\bstar~\alpha \big)$ and append to $\cC$
  \EndFor
  \For{$U_{i,k} \in \bU_i$}
  \State append $\big( \cP_{U_k} $ is a distribution$\big)$ to $\cC$
  \EndFor
  \vspace{1ex}
  \Statex \textit{Optimize}     \\
  \Return optimize $\cT$ subject to $\cC$
\end{algorithmic}
\end{algorithm}

\subsection{Optimizing the Polynomial Program}

Algorithm~\ref{alg:esharp_bound} provides a  step-by-step description of the $\varepsilon$-sharp bounding procedure. For ease of reference, we duplicate the Section~\ref{sec:algo} discussion of the algorithm's various components here.

Algorithm~\ref{alg:esharp_bound} takes as inputs the
polynomialized objective function $\cT(\bp)$ and constraint set $\cC(\bp)$,
obtained from Algorithm~\ref{alg:polynomial-program-simple}. It then evaluates a range of models, or points $\bp$ in the model space $\cP$ for which $\cC(\bp)$ is satisfied. It seeks to identify extreme values of $\cT(\bp)$ within this subspace. It also
accepts two parameters: $\epsilon^{\text{thresh}}$, a stopping threshold for the
looseness factor stopping, and $\theta^{\text{thresh}}$, a stopping threshold
for width of the bounds. The algorithm returns two types of information: the
upper and lower bounds for the causal program, and the worst-case looseness
factor $\varepsilon$.

Primal bounds are denoted $\underline{P}$ and $\overline{P}$, adopting the
convention that underlines refer to objects used for minimization and overlines
for maximization. These indicate the extreme values of the target estimand in
any admissible model---that is, satisfying $\cC(\bp)$---that has been located so
far. These are initialized at $+\infty$ and $-\infty$, respectively, indicating
that no admissible models have been found yet. As optimization proceeds, the
primal bounds improve as new, more extreme admissible models are found. We refer
to $[\underline{P}, \overline{P}]$ as the \textit{inner bounds}: the unknown
sharp bounds must at least contain these points, which correspond to models that
are observationally indistinguishable from the true DGP.

Dual optimization begins by partitioning the parameter space into
branches, proceeding separately for the lower and upper bound and
respectively producing partitions $\underline{\cB}_b$ and $\overline{\cB}_b$. At
initialization, these consist of a single branch spanning the entire
parameter space; each branch is then recursively divided. The lower and upper
parts of the dual envelope, or outer envelope, are denoted $\underline{\cD}$ and
$\overline{\cD}$. These are piecewise linear functions, with pieces
corresponding to the branching partitions, that are \textit{relaxations} of the
true objective function, $\cT(\bp)$, from below and above.  These relaxations are made to ensure they will always
contain the entire objective function at all points in the parameter
space. Within branch $b$,
the value $\min \{ \underline{D}_b(\bp) : \bp \in \overline{\cB}_b \}$ indicates
the lowest value attained by the lower envelope; thus, $\underline{\cT} = \min_b
\l\{ \min \{ \underline{D}_b(\bp) : \bp \in \overline{\cB}_b \} \r\}$ represents
the lowest value attained by the lower envelope anywhere in the parameter
space. Conversely, $\overline{\cT} = \max_b \l\{ \max \{ \overline{D}_b(\bp) :
\bp \in \overline{\cB}_b \} \r\}$ represents the highest value of the upper
envelope. These extreme points on
the dual envelope, $[\underline{\cT}, \overline{\cT}]$, define the dual (outer) bounds. These are the
reported causal bounds; whatever the true
sharp bounds, they must lie inside the dual
bounds, even if the algorithm has not run to completion. We let $\theta$ equal the bound width, or the
difference between the upper and lower dual bounds, and we define the worst-case
looseness factor $\varepsilon$ as the slack (the difference in dual and
primal bound widths) divided by the primal bound width.

The algorithm heuristically selects branches in the model space that appear
promising, and refines primal and dual bounds in turn. It first
searches within the branch for an admissible model; if found, and if the associated causal estimand
is more extreme than those previously encountered, it is stored as a new primal bound. Whatever the true
nonparametric sharp bounds, they must lie outside the primal
bounds because the true bounds must contain the extreme models that define the primal bounds. Then, it
divides the branch into sub-branches and refines the dual envelope by tightening
the piecewise linear outer-approximation. The algorithm
continuously prunes branches of $\underline{\cB}_b$ and $\overline{\cB}_b$ that
are inconsistent with specified constraints; it also continuously
branches and refines the bounds while $\theta$ and $\varepsilon$ exceed
specified thresholds. 

\begin{spacing}{0.8}
\begin{breakablealgorithm}
 \caption{~Computing $\varepsilon$-sharp Bounds}
 \label{alg:esharp_bound}
\vspace{1ex}
\small
\begin{tabular}{ll}
  \textbf{Input:} & target $\cT(\bp)$ and constraint relations $\cC(\bp)$ in parameters $\bp$, \\
  & stopping thresholds $\varepsilon^{\text{thresh}}$ and $\theta^{\text{thresh}}$ \\
  \textbf{Output:} & lower bound $\underline{\cT}$, upper bound $\overline{\cT}$, maximum looseness factor $\varepsilon$
\end{tabular}
\begin{algorithmic}[1]
  \vspace{1ex}
  \Statex \textit{Initialization}  
  \State branches of parameter space: indexed partitions $\underline{\cB} \leftarrow \l\langle [0, 1]^{\hash\{\bp\}} \r\rangle$, $\overline{\cB} \leftarrow \l\langle [0, 1]^{\hash\{\bp\}} \r\rangle$
  \State dual (outer) bounds: indexed families of functions $\underline{\cD} \leftarrow \langle \bp \mapsto -\infty \rangle$, $\overline{\cD} \leftarrow \langle \bp \mapsto +\infty \rangle$
  \State primal (inner) bounds: $\underline{P}=+\infty$ and $\overline{P}=-\infty$
  \State bounds width: $\theta = +\infty$
  \State bounds looseness factor $\varepsilon = +\infty$
  \vspace{1ex}
  \Statex \textit{Spatial branch and bound}  
  \While {$\varepsilon > \varepsilon^{\text{thresh}}$ \textbf{and} $\theta > \theta^{\text{thresh}}$}
  \For {extremum in min, max}
  \vspace{1ex}
  \Statex \hspace*{\algorithmicindent}\hspace*{\algorithmicindent}\textit{Select direction}
  \If{extremum is min}
  \State set $^\ast \leftarrow \underline{\vphantom{X}~~}$ and $\bstar~\leftarrow~\le$
  \ElsIf{extremum is max}
  \State set $^\ast \leftarrow \overline{\vphantom{X}~~}$ and $\bstar~\leftarrow~\ge$
  \EndIf
  %% primal
  \vspace{1ex}
  \Statex \hspace*{\algorithmicindent}\hspace*{\algorithmicindent}\textit{Primal refinement}
  \State continue search for local extremum of $\cT(\bp)$ s.t. $\cC(\bp)$ is satisfied
  \If{feasible point is found \textbf{and} $\cT(\bp)~\bstar~P^\ast$}
  \State update primal bound $P^\ast \leftarrow \cT(\bp)$
  \EndIf
  %% dual
  \vspace{1ex}
  \Statex \hspace*{\algorithmicindent}\hspace*{\algorithmicindent}\textit{Dual refinement}
  \State select outermost branch $b = \argext_{b'}\l\{ {\rm extremum} \{ \cD^\ast_{b'}(\bp) : \bp \in \cB^\ast_{b'} \} \r\}$
  \State pop $\cB^\ast_{b}$ from $\cB^\ast$ and subpartition it, pop $\cD^\ast_{b}(\bp)$ from $\cD^\ast$
  \ForEach {subpartition $\cB^\ast_{b'}$ \textbf{in} $\cB^\ast_{b}$}
  \State push new branch $\cB^\ast_{b'}$ into $\cB^\ast$
  \State find linear function $\cD^\ast_{b'}$
  s.t. $\cD^\ast_{b'}(\bp)~\bstar~\cT(\bp)$ for all $\bp \in \cB_{b'}^\ast$
  \State push linear programming relaxation $\cD^\ast_{b'}$ into $\underline{\cD}$  
  \EndFor
  %% cleanup
  \vspace{1ex}
  \Statex \hspace*{\algorithmicindent}\hspace*{\algorithmicindent}\textit{Prune branches that cannot widen bounds}
  \ForEach {$b$ \textbf{in} $1, \ldots, |\cB^\ast|$}
  \If{$ P^\ast ~\bstar~ {\rm extremum} \{ \cD^\ast_{b}(\bp) : \bp \in \cB^\ast_{b} \} $ \textbf{or} $\cC(\bp)=\rm{False}$ for all $\bp \in \cB^\ast_{b}$}
  \State pop $\cB^\ast_{b}$ from $\cB^\ast$, pop $\cD^\ast_{b}$ from $\cD^\ast$
  \EndIf
  \EndFor
  \vspace{1ex}
  \EndFor
  %$ update
  \vspace{1ex}
  \Statex \hspace*{\algorithmicindent}\textit{Check progress}
  \State $\underline{\cT} \leftarrow \min_b \l\{ \min \{ \underline{D}_b(\bp) : \bp \in \underline{\cB}_b \} \r\}$, $\overline{\cT} \leftarrow \max_b \l\{ \max \{ \overline{D}_b(\bp) : \bp \in \overline{\cB}_b \} \r\}$
  \State $\theta \leftarrow \overline{\cT} - \underline{\cT}$
  \State $\varepsilon \leftarrow \theta / (\overline{P} - \underline{P}) - 1$
  \EndWhile \\
  \vspace{1ex}
 \Return $\underline{\cT}$, $\overline{\cT}$, $\varepsilon$
 \end{algorithmic}
\end{breakablealgorithm}  
\end{spacing}

\section{Proofs}
\label{app:nonrestrictive}

\subsection*{Proof of Proposition \ref{prop:finite-u}.}
\begin{proof}
    We adapt the proof of \cite{wolfe2021cardinalities} to account for
    counterfactuals as follows. First, we define \emph{one-step-ahead}
    counterfactuals, $V_{i,j}\big( \bpa(V_{i,j})=\ba \big)$, to be those where all main parents of a variable are subject
    to intervention $\bpa(V_{i,j})=\ba$. Next, we note that all other counterfactuals and factuals
    in the full data law are deterministic functions of one-step-ahead variables, after fixing $\bU_i$.
    Therefore it is sufficient to reason about only one-step-ahead variables; intervention on other variables is irrelevant to the full data law.
    
    Because the likelihoods of multi-district graphs factorize as the likelihoods of the districts after intervention on their parents \citep{richardson2017nested}, we can consider single-district graphs without loss of generality. In multi-district graphs, the bound obtained below can be applied within each district.
    
    Each main variable $V_{i,j}$ has $|\cS( \bpa(V_{i,j}) )|$ one-step-ahead counterfactuals, corresponding to possible manipulations of its parents. Each one-step-ahead counterfactual $V_{i,j}\big( \bpa(V_{i,j})=\ba \big)$ has a cardinality equal to those of the
    corresponding main variable $| \cS(V_{i,j}) |$.  Therefore, the collection of a single variable's one-step-ahead counterfactuals $\l\langle  V_{i,j}\big( \bpa(V_{i,j})=\ba \big),V_{i,j}\big( \bpa(V_{i,j})=\ba' \big), \ldots \r\rangle$ can take on $|\cS(V_{i,j})|^{|\cS( \bpa(V_{i,j}) )|}$ possible values, and there are $d \equiv \prod_{V_{i,j}
      \in \bV_i} |\cS(V_{i,j})|^{|\cS( \bpa(V_{i,j}) )|} $ values that the full collection of all
    one-step-ahead variables can take. Any model over this full collection must be a subset of the $d
    - 1$ simplex. We let $\bV\big(\bpa(\bV)\big)$ denote the collection of one-step-ahead
    variables. 
    
    Suppose the disturbances $\bU_i$ are enumerated as $\{U_{i,1}, \dots, U_{i,K}\}$. We
    will now show that each $U_{i,k}$ can be assumed to be discrete without altering the model for $\bV\big(\bpa(\bV)\big)$
    and therefore the full data law. First, for each
    value $u_k$ in the domain of $U_{i,k}$, we define the distribution
    $P_{u_k}\Big(\bV\big(\bpa(\bV)\big)\Big) = \int_{\bu_{\setminus k}} P\Big(\bV\big(\bpa(\bV)\big) \mid \bu_{\setminus k}, u_k \Big) P(\bu_{\setminus k})$, where $\bu_{\setminus k}$ denotes all disturbances other than $u_k$. This fixes $U_{i,k}$ at the value $u_k$, modifying the distribution over $\bV\big(\bpa(\bV)\big)$.
    
    We now make two observations. First, the
    model for $\bV\big(\bpa(\bV)\big)$ contains $P_{u_k}$ for any $u_k$, because $U_{i,k}$ is not restricted by the model and is therefore permitted
    to have a point-mass distribution at $u_k$. Second, the expected value of
    $P_{u_k}$ with respect to $U_{i,k}$ recovers the original marginal distribution $P\Big(\bV\big(\bpa(\bV)\big)\Big)$,
    which is therefore in the convex hull of the set of distributions $\cS( P_{u_k} ) \equiv
    \{ P_{u_k} \mid u_k \in \cS(U_{i,k})\}$.
    
    \citeauthor{carathodory1907convex}'s Theorem (\citeyear{carathodory1907convex}) states that for any
    point $P$ in the convex hull of a set $\cS$ in a space of dimension
    $d - 1$, there exists a set of $d - 1$ points $\{P_{u_{k_1}}, \dots,
    P_{u_{k_{d - 1}}}\}$ and weights $\{w_1, \dots, w_{d-1}\}$ such that $P =
    \sum_{\ell = 1}^{d-1} w_\ell P_{u_{i_\ell}}$. It then follows directly that any
    distribution in the marginal model over $\bV\big(\bpa(\bV)\big)$ when latent variables
    have unrestricted cardinality is also in the marginal model over $\bV\big(\bpa(\bV)\big)$
    when latent variables have cardinality restricted to $\prod_{V_{i,j}
      \in \bV_i} |\cS(V_{i,j})|^{|\cS( \bpa(V_{i,j}) )|} - 1$ or higher. 
\end{proof}

\subsection*{Proof of Proposition \ref{prop:polynomialization}}
\begin{proof}
  Using the approach developed in \cite{evans2018margins} and generalized to
  arbitrary graphs in \cite{wolfe2021cardinalities}, we can obtain a functional model
  that is non-restrictive of the causal model of $\cG$ over observed variables.
  In such a model, each $V_{i,\ell}({\bf a}_\ell)$ is determined by by values of
  the disturbances $\bU_i$. By assumption, $\cG$ is in canonical form, rendering
  all disturbances marginally independent. The proposition then follows from
  standard probability calculus.
\end{proof}

\subsection*{Proof of Proposition \ref{prop:relevant-latents}}
\begin{proof}
  Under the conditions specified, no element in $\mathcal C$ involves a function of
  $U_{i,k}$. It follows that whether the disturbances lead to $\mathcal C$ is not a function of the value of $U_{i,k}$. As a result, a sum over all parameters of the
  distribution of $U_{i,k}$ can be factored out of the product in Equation
  \ref{eq:polynomial}. By the definition of probability distributions, this sum will be equal to 1, rendering the
  parameters irrelevant to the polynomial. 
\end{proof}

\subsection*{Proof of Proposition \ref{prop:constraint-complexity}}
\begin{proof}
  Each of the nested Markov parameters corresponds to the probability that
  random variables in a single district take certain values after an intervention
  on parents of the district. It follows from Proposition
  \ref{prop:relevant-latents} that no disturbances outside the district
  corresponding to the nested Markov parameter will appear in the
  polynomialization of that parameter. From this, it then follows that no disturbances
  in different districts will interact in constraints corresponding to nested
  Markov parameters. By Proposition \ref{prop:polynomialization}, the degree of
  a polynomialization of the probability of the event is at most the number of
  relevant disturbances.
\end{proof}

\section{Uncertainty}
\label{app:uncertainty}

In this appendix, we provide details on our approach to quantifying the uncertainty of bounds based on estimated empirical inputs, $\hat{\bE} = \l[ \hat{E}_\ell \r]$. Recall that the \textit{estimated bounds} are obtained from a polynomial program using equality constraints of the form\\ ${\rm polynomialize}\big( g_\ell(\cP_{\bV}) = \hat{E}_\ell \big)$, which is equivalent to $\text{polynomial-fractionalize}\big( g_\ell(\cP_{\bV}) \big) = \hat{E}_\ell$. Here, $\hat{E}_\ell$ is the noisily estimated empirical quantity and $\text{polynomial-fractionalize}\big( g_\ell(\cP_{\bV}) \big)$ is the reexpression of that same quantity in terms of principal strata sizes. At a high level, we will proceed by constructing confidence regions ${\rm
  CR}_\alpha(\hat{\bE})$ such that $\Pr\l( \bE \in {\rm CR}_\alpha(\hat{\bE})
\r) \ge \alpha$. To obtain \textit{confidence bounds}, we then replace empirical equality constraints with a looser version that accounts for sampling variation, of the form $\text{polynomial-fractionalize}\big( g_\ell(\cP_{\bV}) \big) \in {\rm CR}_\alpha(\hat{\bE})$.

Observe that because the main variables are discrete, $\hat{\bE}$ is a realization of a multinomial proportion. In what follows, we will assume that empirical evidence arises from a single multinomial distribution, such as a single-world marginal distribution; if multiple independent sets of empirical evidence about differing quantities are available, the procedure generalizes straightforwardly by repeating the procedure within each set and combining the results appropriately.

Based on this idea, we examine two methods for constructing ${\rm CR}_\alpha(\hat{\bE})$. Drawing on \citet{malloy2020optimal}, we first consider a ``Bernoulli-KL'' approach that constructs separate confidence regions for each observable atomic event, $\Pr(\bV_i = \bv)$, treating it as a ``success'' in a Bernoulli distribution. The approach rotates through all possible $\bv$ and combines the event-specific regions using a result on the Kullback-Leibler divergence of sampling distributions to the underlying population distribution. The Bernoulli-KL method produces a confidence region for single-world distributions that is guaranteed to have conservative coverage for the multinomial proportion in finite samples. 

Let $k \in \{1, \ldots, K\}$ index possible atomic events, and denote the probability of the $k$-th event as $p_k = \Pr(\bV_i = \bv_k)$. Empirical frequencies are denoted $\hat{p}_k$. For the Bernoulli-KL method, we will develop a confidence region of the form ${\rm CR}_\alpha(\hat\bE) = \bigcap_{k=1}^K \l[\underline{p}_k, \overline{p}_k\r]$, noting that each $p_k$ can be polynomialized. A visualization of the resulting region is given in Figure~\ref{fig:uncertainty}(b).

We now describe how $\underline{p}_k$ and $\overline{p}_k$ can be calculated to ensure that $\Pr(\bE \in {\rm CR}_\alpha(\hat\bE)) \ge \alpha$. At a high level, we will do so by analyzing each of the $K$ observable events as a Bernoulli distribution. Taking each $\hat{p}_k$ estimate as given, we identify regions of the unknown $p_k$ from which the observed $\hat{p}_k$ diverge substantially. Equation 11 of \citet{malloy2020optimal} provides bounds on the sampling probability of observing ${\rm KL}\l( [1 - \hat{p}_k, \hat{p}_k], [1 - p_k, p_k] \r)$ in excess of some threshold, where ${\rm KL}\big( [1 - \hat{p}_k, \hat{p}_k], [1 - p_k, p_k] \big) = \hat{p}_k \log \frac{\hat{p}_k}{p_k} + (1 - \hat{p}_k) \log \frac{1 - \hat{p}_k}{1 - p_k}$.

In turn, these bounds imply regions of $p_k$ that can be conservatively rejected. Let $\underline{p}_k$ be given by the solution to ${\rm KL}\big( [1 - \hat{p}_k, \hat{p}_k], [1 - \underline{p}_k, \underline{p}_k] \big) = \frac{1}{N} \log \frac{2 K}{1 - \alpha}$ subject to $\underline{p}_k \in [0, \hat{p}_k]$. Similarly, let $\overline{p}_k$ be given by ${\rm KL}\big( [1 - \hat{p}_k, \hat{p}_k], [1 - \overline{p}_k, \overline{p}_k] \big) = \frac{1}{N} \log \frac{2 K}{1 - \alpha}$ subject to $\overline{p}_k \in [\hat{p}_k, 1]$. It can be seen from \citet{malloy2020optimal} that when constructing $\underline{p}_k$ and $\overline{p}_k$ in this way, $\Pr\l( \bigcap_{k=1}^K p_k \in [\underline{p}_k, \overline{p}_k] \r) \ge \alpha$ over repeated samples.

Our second approach uses an asymptotic confidence region based on the multivariate Gaussian limiting distribution of the multinomial proportion,  $\mathcal{N}\l( \bp, {\rm diag}(\bp) - \bp \bp^\top  \r)$ \citep{bienayme1838memoire}. Because the multinomial proportion must sum to unity, this distribution is degenerate, and it is often more convenient to work with its first $K-1$ elements, $\bp_{\setminus K}$. We construct the asymptotic confidence region as $(\hat{\bp}_{\setminus K} - \bp_{\setminus K})^\top  \l( {\rm diag}(\hat{\bp}_{\setminus K}) - \hat{\bp}_{\setminus K} \hat{\bp}_{\setminus K}^\top \r)^{-1} (\hat{\bp}_{\setminus K} - \bp_{\setminus K}) \le z$, where $z$ is an appropriate critical value of the $\chi^2$ distribution. A visualization of the resulting region is given in Figure~\ref{fig:uncertainty}(c). As before, each element in $\bp$ is polynomializable, leading to a single confidence constraint that can be straightforwardly incorporated into the optimization routine.

\begin{figure}
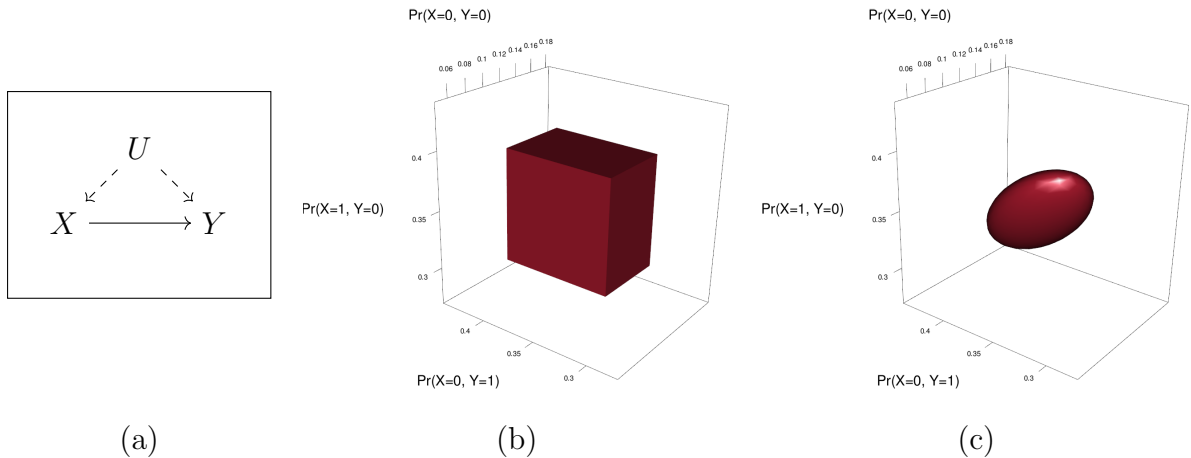

    \caption{ \textbf{Polynomial confidence regions in a binary graph.} Panel (a) presents a causal graph in which binary $X$ causes binary $Y$, but both are confounded by an unobserved $U$. $N=1,000$ observations are drawn from this DGP, producing an empirical distribution with proportions $\frac{1}{N} \sum_{i=1}^N \bbone(X_i=x, Y_i=y)$. Panels (b--c) depict confidence regions for $\Pr(X_i=0, Y_i=0)$, $\Pr(X_i=0, Y_i=1)$, and $\Pr(X_i=1, Y_i=0)$; the final category, $\Pr(X_i=1, Y_i=1)$ (not depicted), must sum to unity. Panel (b) shows the Bernoulli-KL confidence region, which is conservative in finite samples and can be polynomialized as a set of linear inequalities. Panel (c) shows the Gaussian confidence region, which is asymptotically valid and can be polynomialized as a single convex quadratic inequality.}
  \label{fig:uncertainty2}
  \centering
%  \vspace{2ex}
  \begin{tabular}{ccc}
 \begin{tikzpicture}%[node distance=.5 cm and .5 cm]
 \node (U) at (3,1) {$U$};
 \node (X) at (2,0) {$X$};
 \node (Y) at (4,0) {$Y$};
 \path[dashed] (U) edge [->](X);
 \path[dashed] (U) edge [->](Y);
 \path (X) edge [->](Y);
\draw (1.25,-1) rectangle (4.75,1.75);
\useasboundingbox (1.25, -2.5) rectangle (4.75,1.75);
 \end{tikzpicture}
    &
    \includegraphics[width=.35\textwidth, trim={300 50 585 50}, clip]{figures/ci_box.png}
    &
    \includegraphics[width=.35\textwidth, trim={300 50 585 50}, clip]{figures/ci_ellipse.png}
    \\
    (a)
    &
    (b)
    &
    (c)
  \end{tabular}
\end{figure}

For ease of reference, we duplicate  Figure~\ref{fig:uncertainty} in Figure~\ref{fig:uncertainty2}, below. This figure depicts these regions visually for a simple two-node graph, shown in Figure~\ref{fig:uncertainty2}(a). The resulting Bernoulli-KL and Gaussian confidence regions are depicted in Figure~\ref{fig:uncertainty2}(b--c). 

Finally, we describe how arbitrary confidence regions, such as the optimal level-set regions of \citet{malloy2020optimal} or the exact finite-sample regions of \citet{kuchibhotla2021hulc}, can be polynomialized. At a high level, the proposed method uses a circumscribing polytope, adding faces along the region's principal axes until the desired tightness is achieved. 

One possible approach to doing so is to enumerate candidate $\bp$ along a fine grid, assess each candidate for membership in the confidence region, and compute the convex hull of the
non-rejected points. This procedure produces a system of linear inequalities describing the
hull facets. However, it is infeasible for even moderately sized problems, as
the time complexity of hull construction can grow exponentially in
the dimension of the space, $K$ \citep{ottmann1995enumerating}. Our approach builds on
this basic intuition of circumscribing a complex confidence region with a larger,
more tractable polytope. We compute the principal components of the non-rejected
points, then identify the two extreme non-rejected points along each axis. Each principal axis is the normal vector for two boundary planes, and each
extreme point along that axis defines an boundary plane offset. By repeating
this procedure along each principal axis, we obtain a circumscribing
  confidence region, a parallelepiped that contains the KL
confidence region. The gap between the two confidence regions can be rapidly
approximated by using number of grid points that lie in the inscribing region but
not the original confidence region. By slicing the simplex along additional directions, such as
convex combinations of principal axes, this gap can be tightened to arbitrary
precision. The resulting polytope defines a system of linear inequalities that can then be incorporated into the
polynomial program.

\section{Details of Simulated Models}\label{si:sim_models}

In this section, we detail all models presented in Section~\ref{sec:sims}. For simplicity, all main variables in these models are binary. Simulation parameters are described in terms of principal strata. Principal strata can take one of three forms, depending on the number of parents of the relevant variable. Below, we provide compact notation for referring to these principal strata. Subsequent sections report strata probabilities for each simulation, including joint distributions over strata for multiple variables where confounding exists.
\begin{enumerate}
    \item \textbf{Variables with no parents, which have two strata.} Consider a hypothetical variable $X_i$ with no parents, as in Figure~\ref{fig:common}(a). We use $x_0$ to denote units with $X_i(\varnothing) = 0$ and $x_1$ to denote $X_i(\varnothing) = 1$. 
    \item \textbf{Variables with a single parent, which have four strata.} Consider a hypothetical variable $Y_i$ influenced by parent $X_i$, also depicted in Figure~\ref{fig:common}(a). For compactness, we adopt the convention that counterfactual manipulations of parent variables are presented in the form $y_{Y_i(X_i=0),Y_i(X_i=1)}$. For example, (i) we use $y_{00}$ to denote ``never takers'' with example, $Y_i(X_i=0)=0$ and $Y_i(X_i=1)=0$. Similarly, (ii) $y_{01}$ denotes ``compliers'' with $Y_i(X_i=0)=0$ and $Y_i(X_i=1)=1$, (iii) $y_{10}$ denotes ``defiers'' with $Y_i(X_i=0)=1$ and $Y_i(X_i=1)=0$, and $y_{11}$ denotes ``always takers'' with $Y_i(X_i=0)=1$ and $Y_i(X_i=1)=1$.
    \item \textbf{Variables with two parents, which have sixteen strata.} Consider a hypothetical variable $Y_i$ influenced by parents $Z_i$ and $X_i$, as in Figure~\ref{fig:iv}(a). Extending the convention described above, we denote these in compact forms ranging from $y_{0000}$ to $y_{1111}$. Specific definitions are provided in Table \ref{twovarcanmodel}.
  
  \begin{table}[b]
  \centering
  \caption{\textbf{Principal strata for a variable $Y_i$ with two parents, $Z_i$ and $X_i$.} Each row corresponds to a strata, with compact names given in the first column. For each strata, counterfactual values of $Y_i$ are given in subsequent columns.}
  \label{twovarcanmodel}
\begin{tabular}{lrrrr}
\hline\hline
        & \multicolumn{1}{l}{$Y_i(Z_i=0,X_i=0)$} & \multicolumn{1}{l}{$Y_i(Z_i=0,X_i=1)$} & \multicolumn{1}{l}{$Y_i(Z_i=1,X_i=0)$} & \multicolumn{1}{l}{$Y_i(Z_i=1,X_i=1)$)} \\
        \hline
$y_{0000}$ & $0$                            & $0$                            & $0$                            & $0$                            \\
$y_{1000}$ & $1$                            & $0$                            & $0$                            & $0$                            \\
$y_{0100}$ & $0$                            & $1$                            & $0$                            & $0$                            \\
$y_{1100}$ & $1$                            & $1$                            & $0$                            & $0$                            \\
$y_{0010}$ & $0$                            & $0$                            & $1$                            & $0$                            \\
$y_{1010}$ & $1$                            & $0$                            & $1$                            & $0$                            \\
$y_{0110}$ & $0$                            & $1$                            & $1$                            & $0$                            \\
$y_{1110}$ & $1$                            & $1$                            & $1$                            & $0$                            \\
$y_{0001}$ & $0$                            & $0$                            & $0$                            & $1$                            \\
$y_{1001}$ & $1$                            & $0$                            & $0$                            & $1$                            \\
$y_{0101}$ & $0$                            & $1$                            & $0$                            & $1$                            \\
$y_{1101}$ & $1$                            & $1$                            & $0$                            & $1$                            \\
$y_{0011}$ & $0$                            & $0$                            & $1$                            & $1$                            \\
$y_{1011}$ & $1$                            & $0$                            & $1$                            & $1$                            \\
$y_{0111}$ & $0$                            & $1$                            & $1$                            & $1$                            \\
$y_{1111}$ & $1$                            & $1$                            & $1$                            & $1$                           
\end{tabular}
\end{table}
    
\end{enumerate}

\subsection{Noncompliance Simulation}

In this section, we describe the DGP for our noncompliance simulation analyzed in Section~\ref{sec:iv_sim}. The DGP follows the model of Figure~\ref{fig:iv}(b), reproduced below for ease of reference. Simulation parameters are reported in terms of the joint distribution over principal strata.

\begin{figure}[htb]
 \centering
  \caption{\textbf{DGP with noncompliance.}}
 \label{fig:ivappendix}
%\vspace{3ex}
\begin{minipage}[b]{.32\textwidth}
 \centering
 \begin{tikzpicture}%[node distance=.5 cm and .5 cm]
 \node (U) at (3,1) {$U$} ;
 \node (Z) at (0,0) {$Z$};
 \node (X) at (2,0) {$X$};
 \node (Y) at (4,0) {$Y$};
 \path[dashed] (U) edge [->](X);
 \path[dashed] (U) edge [->](Y);
 \path (Z) edge [->](X);
 \path (X) edge [->](Y);
\draw (-.5,-1) rectangle (4.5,1.5);
 \end{tikzpicture}
 \end{minipage}
\end{figure}
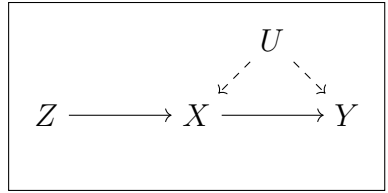

\vspace{0.5cm}

\begin{center}
Strata for $Z$:\\[1ex]

\begin{tabular}{l|r}
\hline
$z_{0}$ & $0.649335$\\
\hline
$z_{1}$ & $0.350665$\\
\hline
\end{tabular}
\end{center}

\vspace{0.5cm}

\begin{center}
Strata for $X$ and $Y$:\\[1ex]

\begin{tabular}{l|r|r|r|r}
\hline
  & $y_{00}$ & $y_{10}$ & $y_{01}$ & $y_{11}$\\
\hline
$x_{00}$ & $0.000757$ & $0.013034$ & $0.006125$ & $0.002606$\\
\hline
$x_{10}$ & $0.004541$ & $0.074105$ & $0.034526$ & $0.014387$\\
\hline
$x_{01}$ & $0.026040$ & $0.418847$ & $0.195419$ & $0.082264$\\
\hline
$x_{11}$ & $0.004534$ & $0.073950$ & $0.034123$ & $0.014742$\\
\hline
\end{tabular}
\end{center}

\vspace{2cm}

\subsection{Outcome-Based Selection Simulation}

In this section, we describe the DGP for our outcome-based selection simulation, analyzed in Section~\ref{sub:gen_sim} and Figure~\ref{fig:common_results}(a). The DGP follows the model of Figure~\ref{fig:common}(a), reproduced below for ease of reference. Simulation parameters are reported in terms of the joint distribution over principal strata.

\begin{figure}[!h]
  \centering
 \begin{minipage}[b]{.45\textwidth}
 \centering
 \begin{tikzpicture}%[node distance=.5 cm and .5 cm]
 \node (U) at (1.5,1) {$U$};
 \node (X) at (0.5,0) {$X$};
 \node (Y) at (2.5,0) {$Y$};
 \node[draw] (S) at (4.5,0) {$S$};
 \path[dashed] (U) edge [->](X);
 \path[dashed] (U) edge [->](Y);
 \path (X) edge [->](Y);
 \path (Y) edge [->](S);
 \draw [rotate around={0:(0,0)}, -, dotted, red, semithick, rounded corners] (.1,-.4) rectangle (2.9,.4);
\draw (-1.25,-.75) rectangle (5.75,1.75);
%\node (a) at (-.95, -1} {(a)};
%\draw (-1.25,-1.25) rectangle (-.65, -.7);
%\draw (-1.25,-1.25) rectangle (5.75,1.75);
%\useasboundingbox (-1.25,-1.25) rectangle (5.75,1.75);
%\useasboundingbox (-1.25,-1.25) rectangle (5.5,1.5);
 \end{tikzpicture}
 \end{minipage}
 \end{figure}

\begin{center}
Strata for $X$ and $Y$\\[1ex]

\begin{tabular}{l|r|r|r|r}
\hline
  & $y_{00}$ & $y_{10}$ & $y_{01}$ & $y_{11}$\\
\hline
$x_{0}$ & $0.124855$ & $0$ & $0.249647$ & $0.124847$\\
\hline
$x_{1}$ & $0.125375$ & $0$ & $0.249851$ & $0.125425$\\
\hline
\end{tabular}
\end{center}

\vspace{0.5cm}

\begin{center}
Strata for $S$\\[1ex]

\begin{tabular}{l|r}
\hline
$S_{10}$ & $0.50052$\\
\hline
$S_{01}$ & $0.49948$\\
\hline
\end{tabular}
\vspace{1cm}
\end{center}

\subsection{Measurement Error Simulation}

In this section, we describe the DGP for our measurement error simulation, analyzed in Section~\ref{sub:gen_sim} and Figure~\ref{fig:common_results}(b). The DGP follows the model of Figure~\ref{fig:common}(b), reproduced below for ease of reference. Simulation parameters are reported in terms of the joint distribution over principal strata.

\begin{figure}[!h]
 \centering
 \begin{minipage}[b]{.45\textwidth}
 \centering
 \begin{tikzpicture}%[node distance=.5 cm and .5 cm]
 \node (U) at (3,0) {$U$};
 \node (X) at (1,0) {$X$};
 \node (Ya) at (2,1) {$Y$};
 \node (Yb) at (4,1) {$Y^\ast$};
 \path[dashed] (U) edge [->](Ya);
 \path[dashed] (U) edge [->](Yb);
 \path (X) edge [->](Ya);
 \path (Ya) edge [->](Yb);
 \draw [rotate around={0:(0,0)}, -, dotted, red, semithick, rounded corners] (.6,-.4) rectangle (1.4,.4);
 \draw [rotate around={0:(0,0)}, -, dotted, red, semithick, rounded corners] (3.6,.6) rectangle (4.4,1.4);
\draw (-1.25,-.75) rectangle (5.75,1.75);
%\node (b) at (-.95, -1} {(b)};
%\draw (-1.25,-1.25) rectangle (-.65, -.7);
%\draw (-1.25,-1.25) rectangle (5.75,1.75);
%\useasboundingbox (-1.25,-1.25) rectangle (5.75,1.75);
 \end{tikzpicture}
 \end{minipage}\\[1ex]
 \end{figure}

\vspace{0.5cm}

\begin{center}
Strata for $X$\\[1ex]

\begin{tabular}{l|r}
\hline
$x_{0}$ & $0.499442$\\
\hline
$x_{1}$ & $0.500558$\\
\hline
\end{tabular}
\end{center}

\vspace{0.5cm}

\begin{center}
Strata for $Y$ and $Y^*$\\[1ex]

\begin{tabular}{l|r|r|r|r}
\hline
  & $Y^*_{00}$ & $Y^*_{10}$ & $Y^*_{01}$ & $Y^*_{11}$\\
\hline
$y_{00}$ & $0$ & $0.167269$ & $0$ & $0$\\
\hline
$y_{10}$ & $0$ & $0$ & $0$ & $0$\\
\hline
$y_{01}$ & $0$ & $0.165838$ & $0.500388$ & $0$\\
\hline
$y_{11}$ & $0$ & $0.166505$ & $0$ & $0$\\
\hline
\end{tabular}
\end{center}

\vspace{1cm}

\subsection{Outcome Missingness Simulation}

In this section, we describe the DGP for our outcome missingness simulation, analyzed in Section~\ref{sub:gen_sim} and Figure~\ref{fig:common_results}(c). The DGP follows the model of Figure~\ref{fig:common}(c), reproduced below for ease of reference. Simulation parameters are reported in terms of the joint distribution over principal strata.

\begin{figure}[!h]
 \centering
 \begin{minipage}[b]{.45\textwidth}
 \centering
 \begin{tikzpicture}%[node distance=.5 cm and .5 cm]
 \node (X) at (0,0) {$X$};
 \node (Ya) at (1,1) {$Y$};
 \node (R) at (3,1) {$R_Y$};
 \node (Yb) at (5,1) {$Y^\ast$};
 \path (X) edge [->](Ya);
 \path (X) edge [->, bend right=20](R);
 \path (Ya) edge [->](R);
 \path (R) edge [->](Yb);
 \path (Ya) edge [->, bend left=20](Yb);
 \draw [rotate around={0:(0,0)}, -, dotted, red, semithick, rounded corners] (-.4,-.4) rectangle (.4,.4);
 \draw [rotate around={0:(0,0)}, -, dotted, red, semithick, rounded corners] (2.6,.6) rectangle (5.4,1.4);
\draw (-1.25,-.75) rectangle (5.75,1.75);
%\node (c) at (-.95, -1} {(c)};
%\draw (-1.25,-1.25) rectangle (-.65, -.7);
%\draw (-1.25,-1.25) rectangle (5.75,1.75);
%\useasboundingbox (-1.25,-1.25) rectangle (5.75,1.75);
 \end{tikzpicture}
 \end{minipage}
 \end{figure}
 
\vspace{0.5cm}

\begin{center}
Strata for $X$\\[1ex]

\begin{tabular}{l|r}
\hline
$x_{0}$ & $0.499159$\\
\hline
$x_{1}$ & $0.500841$\\
\hline
\end{tabular}
\end{center}
\vspace{0.5cm}

\begin{center}
Strata for $Y$\\[1ex]

\begin{tabular}{l|r}
\hline
$y_{00}$ & $0.166371$\\
\hline
$y_{10}$ & $0$\\
\hline
$y_{01}$ & $0.666851$\\
\hline
$y_{11}$ & $0.166778$\\
\hline
\end{tabular}
\end{center}

\vspace{0.5cm}

\begin{center}
Strata for $R$\\[1ex]

\begin{tabular}{l|r}
\hline
$r_{0000}$ & $0$\\
\hline
$r_{1000}$ & $0$\\
\hline
$r_{0100}$ & $0.250368$\\
\hline
$r_{1100}$ & $0.249910$\\
\hline
$r_{0010}$ & $0$\\
\hline
$r_{1010}$ & $0$\\
\hline
$r_{0110}$ & $0$\\
\hline
$r_{1110}$ & $0$\\
\hline
$r_{0001}$ & $0$\\
\hline
$r_{1001}$ & $0$\\
\hline
$r_{0101}$ & $0.250154$\\
\hline
$r_{1101}$ & $0$\\
\hline
$r_{0011}$ & $0$\\
\hline
$r_{1011}$ & $0$\\
\hline
$r_{0111}$ & $0$\\
\hline
$r_{1111}$ & $0.249568$\\
\hline
\end{tabular}
\end{center}

\vspace{1cm}

\subsection{Joint Missingness Simulation}

In this section, we describe the DGP for our joint missingness simulation, analyzed in Section~\ref{sub:gen_sim} and Figure~\ref{fig:common_results}(d). The DGP follows the model of Figure~\ref{fig:common}(d), reproduced below for ease of reference. Simulation parameters are reported in terms of the joint distribution over principal strata.

\begin{figure}[!h]
 \centering
\begin{minipage}[b]{.45\textwidth}
 \centering
 \begin{tikzpicture}%[node distance=.5 cm and .5 cm]
 \node (Xa) at (0,0) {$X$};
 \node (RX) at (2,0) {$R_X$};
 \node (Xb) at (4,0) {$X^\ast$};
 \node (Ya) at (1,1) {$Y$};
 \node (RY) at (3,1) {$R_Y$};
 \node (Yb) at (5,1) {$Y^\ast$};
 \path (Xa) edge [->](Ya);
 \path (Ya) edge [->](RY);
 \path (RX) edge [->](RY);
 \path (RX) edge [->](Xb);
 \path (Xa) edge [->, bend right=20](Xb);
 \path (RY) edge [->](Yb);
 \path (Ya) edge [->, bend left=20](Yb);
 \draw [rotate around={0:(0,0)}, -, dotted, red, semithick, rounded corners] (1.6,-.4) rectangle (5.4,1.4);
% \draw [rotate around={0:_{0,0}}, -, dotted, red, semithick, rounded corners] _{1.2,-.4) -- (2.6,1.4) -- (5.6,1.4) -- (4.2,-.4) -- cycle;
\draw (-1.25,-.75) rectangle (5.75,1.75);
%\node (d) at (-.95, -1} {(d)};
%\draw (-1.25,-1.25) rectangle (-.65, -.7);
%\draw (-1.25,-1.25) rectangle (5.75,1.75);
%\useasboundingbox (-1.25,-1.25) rectangle (5.75,1.75);
 \end{tikzpicture}
 \end{minipage}
\end{figure}

\begin{center}
Strata for $X$\\[1ex]

\begin{tabular}{l|r}
\hline
$x_{0}$ & $0.43464$\\
\hline
$x_{1}$ & $0.56536$\\
\hline
\end{tabular}
\end{center}
\vspace{0.5cm}

\begin{center}
Strata for $Y$\\[1ex]

\begin{tabular}{l|r}
\hline
$y_{00}$ & $0.485336$\\
\hline
$y_{10}$ & $0.253616$\\
\hline
$y_{01}$ & $0.003768$\\
\hline
$y_{11}$ & $0.257279$\\
\hline
\end{tabular}
\end{center}
\vspace{0.5cm}

\begin{center}
Strata for $R_x$\\[1ex]

\begin{tabular}{l|r}
\hline
$r_{x,0}$ & $0.470201$\\
\hline
$r_{x,1}$ & $0.529798$\\
\hline
\end{tabular}
\end{center}
\vspace{0.5cm}

\begin{center}
Strata for $R_y$\\[1ex]

\begin{tabular}{l|r}
\hline
$r_{y,0000}$ & $0$\\
\hline
$r_{y,1000}$ & $0$\\
\hline
$r_{y,0100}$ & $0.162045$\\
\hline
$r_{y,1100}$ & $0$\\
\hline
$r_{y,0110}$ & $0.177470$\\
\hline
$r_{y,0001}$ & $0.107010$\\
\hline
$r_{y,1001}$ & $0.120311$\\
\hline
$r_{y,0101}$ & $0.255778$\\
\hline
$r_{y,1101}$ & $0.081733$\\
\hline
$r_{y,0011}$ & $0$\\
\hline
$r_{y,1011}$ & $0$\\
\hline
$r_{y,0111}$ & $0.095652$\\
\hline
$r_{y,1111}$ & $0$\\
\end{tabular}
\end{center}
\vspace{0.5cm}

\end{document}